\documentclass[preprint2]{emulateapj}
\usepackage{graphicx}

\def\chandra{{\it Chandra}}
\def\ec{$\eta$~Car}
\def\eca{$\eta$~Car\,A}
\def\ecb{$\eta$~Car\,B}

\def\rxte{{\it RXTE}}
\def\hst{{\it HST}}
\def\xmm{{\it XMM-Newton}}

\def\fuse{{\it FUSE}} 

\def\kms{km~s$^{-1}$}

\def\cm{cm$^{-1}$}
\def\lm{$\lambda$}

\newcommand{\Rsun}{\hbox{$R_\odot$}}

\newcommand{\altion}[2]{\textup{#1}\,\textsc{#2}}

\begin{document}

\title{Eta Carinae across the 2003.5 minimum: Spectroscopic Evidence for Massive  Binary Interactions}

\shorttitle{Eta Car: Spectroscopic Evidence for Massive Binary Interaction}
\shortauthors{Nielsen et al.}

\author{K.~E.~Nielsen\altaffilmark{1,2}, M.~F.~Corcoran\altaffilmark{3,4},  T.~R.~Gull\altaffilmark{2}, D.~J.~Hillier\altaffilmark{5}, K.~Hamaguchi\altaffilmark{3,4},  S.~Ivarsson\altaffilmark{1} and D.~J.~Lindler\altaffilmark{2,6}}

\altaffiltext{1}{Catholic University of America, Washington DC\,20064}

\altaffiltext{2}{NASA Goddard Space Flight Center, Astrophysics Science Division, Code 667,  Greenbelt, MD 20771}

\altaffiltext{3}{NASA Goddard Space Flight Center CRESST, Astrophysics Science Division, Code 662, Greenbelt, MD 20771}

\altaffiltext{4}{Universities Space Research Association, 10211 Wincopin Circle, Suite 500, Columbia, MD 21044}

\altaffiltext{5}{Department of Physics and Astronomy, University of Pittsburgh, Pittsburgh, PA 15260}

\altaffiltext{6}{Sigma Space Corporation, 4801 Forbes Boulevard, Lanham, MD 20706}

\email{nielsen@milkyway.gsfc.nasa.gov}

\begin{abstract}
We have analyzed high spatial, moderate spectral resolution observations of Eta Carinae (\ec) obtained with the Space Telescope Imaging Spectrograph (STIS) from 1998.0 to 2004.3. The data were obtained at discrete times covering an entire 2024 day spectroscopic cycle, with focus on the X-ray/ionization low-state which began in 2003 June. The spectra show prominent P-Cygni lines in \ion{H}{1}, \ion{Fe}{2} and \ion{He}{1} which are complicated by blends and contamination by nebular emission and absorption along the line-of-sight toward the observer. All lines show phase and species dependent variations in emission and absorption. For most of the cycle the \ion{He}{1} emission is blueshifted relative to the \ion{H}{1} and \ion{Fe}{2} P-Cygni emission lines, which are approximately centered at system velocity. The  blueshifted \ion{He}{1} absorption varies in intensity and velocity throughout the 2024 day period. We construct radial velocity curves for the absorption component  of the \ion{He}{1} and \ion{H}{1} lines. The \ion{He}{1} absorption shows significant radial velocity variations throughout the cycle, with a rapid change of over 200 \kms\ near the 2003.5 event. The \ion{H}{1} velocity curve is similar to that of the \ion{He}{1} absorption, though offset in phase and reduced in amplitude.  We interpret the complex line profile variations in \ion{He}{1}, \ion{H}{1} and \ion{Fe}{2} to be a consequence of the dynamic interaction of the dense wind of \eca\ with the less dense, faster wind plus the radiation field of a hot companion star, \ecb. The companion's wind carves out a cavity in \eca's wind, which allows UV flux from \ecb\ to penetrate and photoionize an extended region of \eca's wind. During most of the orbit, \ecb\ and the He$^{+}$ recombination zone are on the near side of \eca, producing blueshifted \ion{He}{1} emission. \ion{He}{1} absorption is formed in the part of the He$^{+}$ zone that intersects the line-of-sight toward \ec. We use the variations seen in \ion{He}{1} and the other P-Cygni lines to constrain the geometry of the orbit and the character of \ecb.  
\end{abstract}


\section{Introduction}
Eta Carinae (\ec) is a superluminous, unstable object which underwent a rapid brightening in the 1840s, accompanied by ejection of a substantial amount of material.  These ejecta now form a structured, bipolar nebula, the Homunculus, which makes direct observations of the star difficult. Hence, the nature of \ec\ has long been debated. Most observations up through the mid 1990s could be interpreted as arising from either a single star or binary system \citep{Davidson97b}. More recent evidence favors the binary star interpretation. Indirect signatures of the companion star are periodic events first discovered as intensity variations in \ion{He}{1} \lm10830 \citep[and references therein]{Damineli96} with a 5.5 year periodicity. The excitation events are accompanied by eclipse-like minima observed with UBV and BVR photometry \citep{vanGenderen03}, in near infrared JHKL photometry \citep{Whitelock04} and in X-ray brightness \citep{bish99, Corcoran05}. The X-ray light-curve as measured with the \textit{Rossi X-ray Timing Explorer} (\rxte) is characterized by a gradual increase in brightness just before a rapid decline to a low-state which lasts $\sim$70 days. \rxte\ observations of two such cycles fixed the orbital period to $2024\pm2$ days (5.54 years), which is in agreement with the period derived from the infrared observations and from the \ion{He}{1} \lm10830 variability. Emission from the Weigelt blobs \citep{Weigelt86} located within 0\farcs3 of \ec\ also shows low-excitation intervals tracing the spectroscopic period, with emission from highly ionized species such as [\ion{Ar}{3}] and [\ion{Fe}{4}] disappearing during the short spectroscopic minimum. \citet{Verner05a} attributed this variation to the visible presence of an additional source of flux, possibly an O or Of/WN7 star with $T_\mathrm{eff}$$\sim$37,000~K,  during the time outside of the minima. \\ 
\indent The X-ray variations, as seen by \rxte, \chandra\ and \xmm, have been modeled as a collision between a massive wind of the primary star (\eca) and a less dense wind of the hot companion (\ecb) in a highly elliptical orbit \citep{bish99,Corcoran01, Hamaguchi07}. Additional evidence for the presence of a  binary component is the detection of \ion{He}{2} \lm4686 which appeared in ground-based spectra of \ec\ \citep{Steiner05, Stahl05} shortly before the minimum, but disappeared at the onset of the minimum. \citet{Gull05c} and \citet{Martin05} confirmed the presence of \ion{He}{2} \lm4686 with high angular resolution \textit{Hubble Space Telescope} (\hst) STIS spectra, and  \cite{Martin05} proposed that the \ion{He}{2} \lm4686 emission is formed in the interface between the stellar winds or possibly arising from photo-ionization by the companion star. \\
\indent The multi-wavelength variations observed in \ec\ have been interpreted as arising from the interaction of the winds of \eca\ and \ecb\ or the impact of the radiation from the hot \ecb\ on the circumstellar material. Recently, \citet{Iping05} reported evidence of the companion through detection with the \textit{Far Ultraviolet Spectroscopic Explorer} (\fuse) of a source of far-UV radiation which disappeared just before the beginning of the 2003.5 event, and reappeared by 2004 March. \\
\indent Binary system models have been discussed  by \citet{Damineli97, bish99, Corcoran01, Falceta05, Soker05}, among others. In these models the far-UV flux from a hot companion alters the ionization state of \eca's wind, while X-rays are produced in a bow-shock due to the collision of the dense wind of \eca\ with the lower density, higher velocity wind of \ecb.  In most of the colliding wind binary models, the low-state in observed X-ray intensity, the brightness in \ion{He}{1} \lm10830 and in the visible-band, IR and radio fluxes, occurs when \ecb\ along with the wind-wind collision region moves behind \eca\ and is occulted by its dense wind. Other interpretations are that the reduction in X-ray flux is primarily due to mass transfer between the stars for a brief period near periastron passage \citep{Soker05} or that the X-ray flux is reduced by a dense accumulation of wind material trailing \ecb\ after periastron passage \citep{Falceta05}. In the \citeauthor{Falceta05} model, the companion star is positioned between the observer and the primary star during the periastron passage. This orientation is difficult to reconcile with the radiation cut-off observed toward the Weigelt blobs during the minimum. The Weigelt blobs are located on the same side of \ec\ as the observer, based on proper motions and observed blueshifted emission \citep{Davidson97, Zethson01b, Nielsen06}. The \citeauthor{Falceta05} orientation is also difficult to reconcile with \hst\ Advanced Camera for Surveys (ACS) imagery \citep{Smith04b}. \\
\indent The \ec\ binary system bears striking similarities to the canonical long-period colliding wind binary, WR~140. WR~140 is a massive system (total system mass $\sim$70$M_{\odot}$) composed of a WC7 +O4-5 pair in a 7.9-year, highly eccentric orbit \citep{Marchenko03}.  The strong wind from the WC7 star dominates the emission from the system, though the O4-5 star is the more massive component and dominates the total luminosity of the system. WR~140 is an X-ray variable, with an X-ray light-curve which rapidly climbs to a maximum of $\sim$5$\times$10$^{33}$ ergs s$^{-1}$ in the 0.5$-$10 keV band, followed by a rapid decline to minimum near periastron passage \citep{Pollock05}.  Like \ec, WR~140 shows infrared \citep{Williams90} and radio \citep{White95} variability correlated with the orbital period.  Like \ec, the infrared brightness of WR~140 peaks near periastron passage, while the radio flux drops precipitously at that time.  Unlike \ec, in WR~140 both stellar components can be directly observed; and unlike \ec, WR~140 is not shrouded by thick circumstellar ejecta.\\
\indent As part of an \hst\ Treasury project\footnote{see \url{http://archive.stsci.edu/prepds/etacar/}}, \ec\ was observed with \hst\ STIS throughout the  2024 day (5.54 year) cycle including the spectroscopic low-state which began on 2003 June 29. The STIS spectra provide the highest spatial resolution spectrometry of the star ever obtained and thus offer the best data-set to decouple stellar changes from changes in the circumstellar nebula. STIS's extended wavelength coverage into the ultraviolet probed for direct evidence of the companion star in the form of excess continuum and potential wind lines characteristic of a hot star. The  optical wind lines in the stellar spectrum are excellent diagnostics of the companion's influence on \eca's wind. Of particular importance are the P-Cygni lines from excited levels in neutral  helium. The population of the excited helium energy states requires high energy photons, and is dependent on the  radiation from the companion star.  \citet{Verner05a} found that radiation from \ecb\ is necessary to produce nebular \ion{He}{1} emission lines observed in the spectra of the Weigelt blobs. \\
\indent This paper describes the characteristics of the \ion{He}{1} lines in the STIS spectra throughout the orbit, and compares their behavior to lines in \ion{H}{1}, \ion{Fe}{2} and [\ion{N}{2}]. Since these lines are  formed under different physical conditions in the \ec\ system, they sample the response of the system to the interactions with the companion.We derive a velocity curve based on the \ion{He}{1} absorption, and show that it is  similar to the expected photospheric radial velocity curve for a star in a highly eccentric orbit. The spectral line analysis is used to constrain the system parameters and indicates that the \ion{He}{1} absorption column samples a spatially complex region that changes continuously throughout the spectroscopic period due to dynamical and radiative interactions between the two stars in the \ec\ system.
    
\begin{deluxetable*}{llccc} 
\tabletypesize{\scriptsize} 
\tablecaption{Data used in this analysis. Observations made with the G430M or G750M grating. \label{t1}} 
\tablewidth{0pt} \tablehead{
\colhead{Proposal id.} &
\colhead{Observation Date} &
\colhead{JD } & 
\colhead{Orbital Phase, $\phi$\tablenotemark{a}}  &
\colhead{Position Angle}\\
\colhead{} &
\colhead{} &
\colhead{(+2,450,000)} & 
\colhead{}  &  
\colhead{(deg)}
}
\startdata
7302\tablenotemark{b}  & 1998 January 1        & 0814 &  0.007 & $-$100 \\ 
                                           & 1998 March 19         & 0891 & 0.045  & \phn$-$28 \\
8036\tablenotemark{c}  & 1998 November 25 & 1142 & 0.169  & \phs133 \\
   			               &  1999 February 21   & 1231 & 0.213  & \phn$-$28 \\
8327\tablenotemark{b}  &  2000 March 13        & 1616 & 0.403  &\phn $-$41\\
8483\tablenotemark{c}  & 2000 March 20         & 1623 & 0.407 &  \phn$-$28\\
8619\tablenotemark{b}  & 2001 April 17 	         & 2017 & 0.601 & \phs\phn22 \\
9083\tablenotemark{b}  & 2001 October 1        & 2183 & 0.683  & \phs165 \\
8619\tablenotemark{b}  & 2001 November 27 & 2240 & 0.712  & $-$131\\
9083\tablenotemark{b}  & 2002 January 19$-$20  & 2294  & 0.738 & \phn$-$82 \\
9337\tablenotemark{b}  & 2002 July 4 & 2460 & 0.820 & \phs\phn69 \\
9420\tablenotemark{b}  & 2002 December 16 & 2664 & 0.921 & $-$115 \\ 
           & 2003 February 12$-$13 & 2683 & 0.930 & \phn$-$57\\   
           & 2003 March 29 & 2727 & 0.952 & \phn$-$28 \\
           & 2003 May 5 & 2764 & 0.970  & \phs\phn27\\  
           & 2003 May 17 & 2776 & 0.976 & \phs\phn38\\
           & 2003 June 1 & 2792  & 0.984 &  \phs\phn62\\
           & 2003 June 22 & 2813 & 0.995 & \phs\phn70 \\
9973\tablenotemark{b}  & 2003 July 5 & 2825 & 1.001 & \phs\phn69 \\
           & 2003 July 29 & 2851 & 1.013  & \phs105\\
           & 2003 September 22 & 2904 & 1.040 & \phs153 \\
           & 2003 November 17 & 2961 & 1.068 & $-$142 \\
           & 2004 March 6 & 3071  & 1.122 & \phn$-$29 \\
\enddata
\tablecomments{All \hst\ STIS spectra used in this analysis are available on \url{http://archive.stsci.edu/prepds/etacar/} or  \url{http://etacar.umn.edu/}}
\tablenotetext{a} {Phase relative to the beginning of the \rxte\ low-state 1997.9604: JD2,450,799.792+2024$\times\phi$ \citep{Corcoran05}}
\tablenotetext{b} {\hst\ GO  Program -- K. Davidson (PI)}
\tablenotetext{c} {STIS GTO  Program -- T.~R. Gull (PI)}
\end{deluxetable*}

\begin{figure}
\centering
   \resizebox{\hsize}{!}{\includegraphics{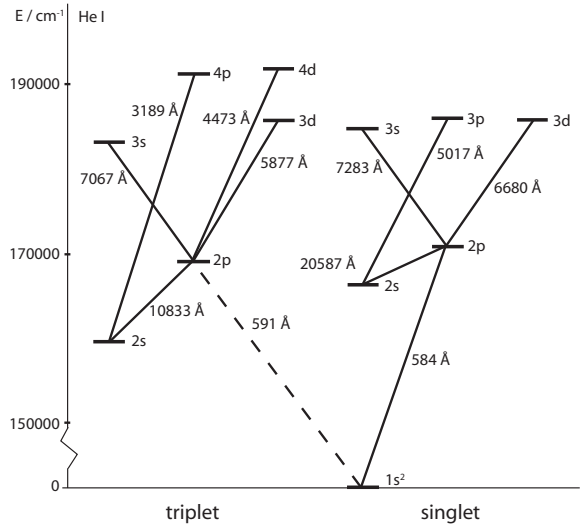}} 
   \caption{Energy level diagram including all \altion{He}{i} transitions used in this analysis. Note, the amount of energy needed to populate the excited states (1 \cm\  = 1.240$\times$10$^{-4}$ eV). All wavelengths are in vacuum.  \label{f1_lab}} 
\end{figure}

\section{Observations}
Spectroscopic observations were conducted with the \hst\ STIS beginning 1998 January~1, just after the X-ray drop, as  seen by \rxte\ \citep{Corcoran05}, and continuing at selected intervals through 2004 March~6. The data set covers more than one full spectroscopic period (5.54 years), with a focus on the 2003.5 event as a part of the \ec\ \hst\ Treasury project. The relatively short time span of the observation (6.2 years) compared to the spectroscopic period, restricts comparisons between long term trends and cyclic variations. To better understand the long term trends we need to have high spectral and spatial resolution spectra with good temporal sampling across the next event, occurring in early 2009.   All observations utilized the 52$''$$\times$0\farcs1 long aperture in combination with the G430M or G750M gratings, yielding a spectral resolving power of R$\sim$6000$-$8000 and a signal-to-noise of 50$-$100 per spectral resolution element. As observations were desired at critical phases of the period, spacecraft solar power restrictions determined the slit position angle. Hence, the  observations were done with a range of position angles. When possible, observations were scheduled with slit orientation at position angle of $-$28$^\circ$ or 152$^\circ$ to include the spatially resolved Weigelt B and D blobs within the long aperture. With superb cooperation with the \hst\ schedulers, this was accomplished several times throughout the 6.2 year observational  interval. \\
\indent The advantage of \hst\ STIS over ground-based instruments is the high angular resolution, which permits exclusion of the bright nebular contributions of scattered starlight and nebular emission lines. While an ideal extraction of two pixels (0\farcs101) would be desirable, reduction issues due to small tilts of the dispersed spectrum with respect to the CCD rows can lead to an artificially induced modulation of the point-source spectrum. A tailored data reduction using interpolation was accomplished with half-pixel sampling by Davidson, Ishibashi and Martin (\url{http://etacar.umn.edu})  minimizing the modulation. We determined that a six half-pixel extraction (0\farcs152) of the online reduced data provided an accurate measure of the stellar spectra while minimizing the nebular contamination.  The 0\farcs152 extraction covers a region corresponding to $\sim$350 AU at the distance of \ec, i.e. much larger than the major axis of the binary orbit (30 AU). The six half-pixel extraction excludes most of the radiation from the Weigelt blobs located 230$-$575 AU from \ec, characterized by narrow line emission. However, even with the 0\farcs152 extraction the wind lines are slightly influenced by some of the strong nebular emission features. The observed spectra used in this paper, with corresponding orbital phase\footnote{All observations in this paper are referenced to the beginning of the \rxte\ low-state 1997.9604: JD2,450,799.792+2024$\times\phi$ \citep{Corcoran05}}, are summarized in Table~\ref{t1}. The complete \hst\ STIS spectra for all phases are available on \url{http://archive.stsci.edu/prepds/etacar/} or  \url{http://etacar.umn.edu/}.   

\begin{figure}
   \centering
   \resizebox{!}{\hsize}{\includegraphics{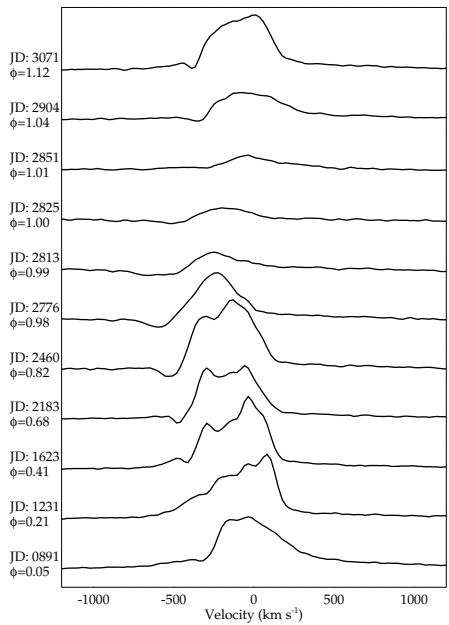}}
  \caption{\altion{He}{i} \lm7067 profiles from 1998.0 to 2004.3. The \altion{He}{i} lines profiles, mostly blueshifted relative to the $-$8 \kms\ system velocity, are asymmetric and vary in intensity and velocity with phase. The 2003 June 23  (Julian Day 2,452,813 $=$ JD: 2813) observation shows greater absorption at higher velocities ($v$$\sim$$-$800 \kms) than any of the other observations. This large  blue-shift is due to a second absorption system at higher velocity. At phase, $\phi$=0.21, the absorption is weak and the velocity measurement is uncertain. \label{f2_lab} }
\end{figure}

\begin{figure}
   \centering
   \resizebox{\hsize}{!}{\includegraphics{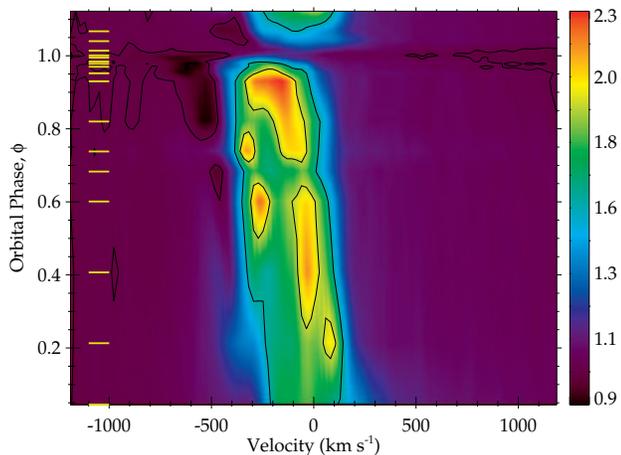}}
   \caption{Surface plot for \altion{He}{i} $\lambda$7067. The line profile's variation with phase is presented with the intensity color-coded from black to red with increasing line emission. All spectra used in the figure has been normalized to the flux at $-$1200 \kms. The observed phase points are marked to the left in the figure. Notice that the two emission components merge just before periastron with a distinct velocity shift at phase, $\phi$=1.0 (2003 June 29).  The absorption, in contrast to that shown by \altion{H}{i}, is most prevalent at $\phi$=0.8$-$1.0. The contours represent the flux level at $-$1200 \kms\ plus 0.70 and 0.85 of maximum line emission.  \label{f3_lab}}
\end{figure}

\begin{figure}
   \centering
 \resizebox{!}{\hsize}{\includegraphics[angle=0]{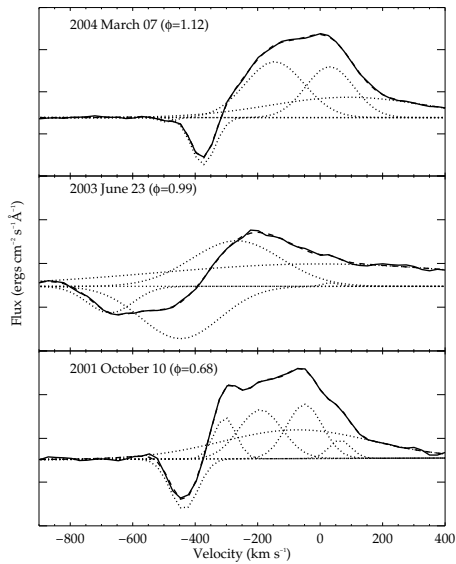}}
  \caption{The \altion{He}{i} lines can be described by a broad P-Cygni wind profile with the superposition of several narrow emission components as exemplified with the \altion{He}{i} $\lambda$7067 profiles and component fits. Solid line: Observed spectrum. Dashed line: Fitted line profile based on a sum of the individual components (dotted lines). \label{f4_lab}}
\end{figure}

\begin{figure}
   \centering
   \resizebox{!}{\hsize}{\includegraphics[angle=0]{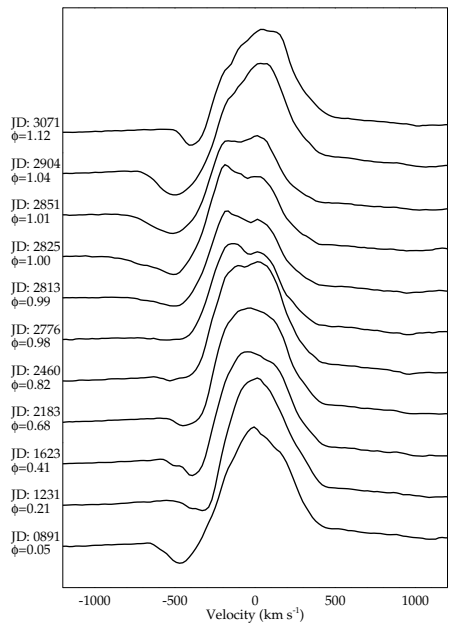}} 
   \caption{\altion{H}{i} \lm4103 variability in intensity and velocity over the spectroscopic period. The emission centroid and emission fluxes are close to the system velocity ($-$8 \kms), and it and the emission fluxes do not generally undergo the dramatic changes observed in \altion{He}{i}. However, significant changes in the emission line radial velocity and the \altion{H}{i} emission line strength occurs near phase, $\phi$=1.0. Changes are observed in the P-Cygni absorption over the entire cycle, with the absorption weakest at $\phi$=0.21$-$0.98. The absorption velocity varies over the period but with a smaller amplitude than observed in \altion{He}{i}. \label{f5_lab}} 
\end{figure}

\begin{figure}
   \centering
   \resizebox{\hsize}{!}{\includegraphics[angle=0]{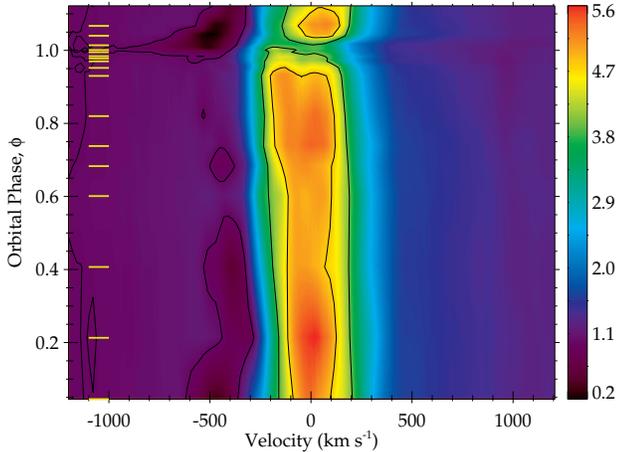}} 
   \caption{Surface plot for \altion{H}{i} $\lambda$4103. The line profile's variation with phase is presented with the intensity color-coded from black to red with increasing line emission. All spectra used in the figure has been normalized to the flux at $-$1200 \kms, and the phase points are marked on the left of the figure. The emission profiles is nearly constant, both in intensity and velocity, throughout most of the period. Significant changes are seen in the emission profile near phase, $\phi$=1.0; there is a reduction in flux and a velocity shift. An abrupt strengthening of the P-Cygni absorption occurs after phase, $\phi$=1.0.  The contours represent the flux level at $-$1200 \kms\ plus 0.70 and 0.85 of maximum line emission. \label{f6_lab}}
\end{figure}

\section{Spectral Analysis} 
The spectrum of \ec\ contains a wealth of information. The spectrum is dominated by wind lines from \eca, in particular the \ion{H}{1} Balmer and Paschen series and numerous lines from the singly ionized iron-group elements. A detailed discussion of the optical spectrum, and its formation, is given by \citet{Hillier01}. The UV wind spectrum is described in \citet{Hillier01, Hillier06}, where the  general characteristics are discussed including the absence of wind lines from \ecb. We concentrate on the orbital phase dependance of four species whose spectra dominate the optical wavelength region: \ion{H}{1} (ionization potential 13.6 eV), \ion{He}{1} (IP 24.6 eV), \ion{Fe}{2} (\ion{Fe}{2} IP 16.2 eV with \ion{Fe}{3} IP 30.7 eV), and [\ion{N}{2}]\ \lm5756 (\ion{N}{1} IP 14.5 eV with \ion{N}{2} IP 29.6 eV).  Due to the large  abundance of hydrogen and  its ionization potential, \ion{H}{1} lines form throughout the stellar wind structure. Iron shows similar behavior to hydrogen but due to its ionization structure, \ion{Fe}{2} lines form in the outer lower excited regions of \eca's wind, with the interior region dominated by more highly ionized iron. Helium is mainly neutral in the environment of the primary star. However, the energy level structure (see Figure~\ref{f1_lab}) is such that high energy photons are required even to populate the lowest excited states. \ion{He}{1} lines represent the highly excited regions of the stellar wind or the bow-shock. The  [\ion{N}{2}] \lm5756 line is a parity forbidden line, and is thus formed in the outer, low density, ionized region. \\
\indent STIS optical spectra of the star show that the excited \ion{He}{1} stellar lines are peculiar relative to lower excitation wind lines: the \ion{He}{1} lines are blueshifted relative to the system velocity ($-$8 \kms, \citealt{Smith04c}) over most of the 5.54 year cycle, unlike the lower excitation lines, whose emission is centered near system velocity. The \ion{He}{1}  P-Cygni lines in the \ec\ spectrum are characterized by a broad asymmetric P-Cygni profiles which often exhibit multiple, discrete peaks. Another unusual characteristic of the \ion{He}{1} lines is the abrupt, large velocity shift towards the system velocity,   exhibited in both emission and absorption, which occurred during the 2003.5 minimum. The variations of the \ion{He}{1} P-Cygni lines in both strength and radial velocity suggest that most of the \ion{He}{1} lines are influenced by the ionizing flux of the companion star \citep{Hillier06}.\\
\indent While the spectrum of \ec\ is rich in wind lines, many lines are not suitable for detailed analysis. A large number of \ion{He}{1} lines are located in the STIS CCD spectral range (1600$-$10,300 \AA), but many, such as the commonly used \ion{He}{1} \lm4473, are either blended or too faint for reliable interpretation. Even with the 0\farcs152 extraction, a few of the  lines are influenced by emission from the surrounding nebula and the Weigelt blobs. The  \ion{He}{1} \lm\lm5877, 6680 lines are slightly affected by narrow, nebular [\ion{Fe}{2}] and [\ion{Ni}{2}] emission lines, respectively. The least contaminated \ion{He}{1} transition in the spectrum, for both absorption and emission, is \ion{He}{1} \lm7067. Hence, we use this spectral line to illustrate the variability in absorption and emission. Unfortunately, the absorption in \ion{He}{1} \lm7067 is often weak and difficult to measure. We have therefore also utilized the absorption portion of \ion{He}{1} \lm5877. The \ion{He}{1} \lm5877 line has the same lower state as \ion{He}{1} \lm7067, but its oscillator strength is almost one order of magnitude  greater and its absorption is correspondingly stronger. The \ion{He}{1} \lm5877 emission is heavily affected by the strong \ion{Na}{1} \lm5892 resonance absorption from the stellar wind, the intervening ejecta and the interstellar medium. The  \ion{He}{1} \lm5017 line is badly blended with \ion{Fe}{2} \lm5020, and consequently little information can be gleaned about the emission profile. However, we can estimate the influence of the blend on the associated absorption using other members of the same multiplet, such as \ion{Fe}{2} \lm5170. We find that the absorption is dominated by the \ion{He}{1} component throughout most of the period but is overwhelmed by \ion{Fe}{2} absorption for a short interval after periastron (Section~\ref{sect_hei_var}). A complete set of lines used in this analysis is presented in Table~\ref{t2}, while a detailed discussion of a subset of the least blended transitions is presented in Section~4.\\
\indent To investigate the strongly phase dependent and complex profile variability we utilized a variety of techniques. To illustrate these we use \ion{He}{1} \lm7067 which shows strong variability at all phases. At least four components can be identified in \ion{He}{1} \lm7067: one absorption component, two narrow emission components which give rise to the bulk of the observed emission, and a much weaker, broad emission component. All these components show significant time variability. \\  
\indent To illustrate the observed spectral variations, we stacked continuum-normalized line plots in velocity space (Figure~\ref{f2_lab}). We also used a color-coded intensity plot to illustrate the variation as a function of orbital phase. We created the color-coded plots by stacking the spectra according to phase, normalizing to the flux at $-$1200 \kms\ relative to the vacuum rest wavelengths of the wind lines, where the continuum is seen to be unaffected by the wind line variations. The flux between the observations has been calculated by interpolation in phase to help identify systematic trends. The normalization procedure is used for the visualizations only, and utilized to emphasize the spectral variations with phase. The systematic change in radial velocity is apparent, as is a discontinuity in the radial velocity of the \ion{He}{1} absorption near phase 1.0 (Figure~\ref{f3_lab}). Quantifying the observed spectral changes is difficult, since both the shape and strength of the line change. In addition, reliable measurements of absolute line fluxes are difficult since the amount of extinction is declining with time. Since 1997, the brightness of the central star, as observed by \hst\ in H$\alpha$, has increased  by over a factor three  \citep{Martin04}. During the minimum, small variations (10$-$20\%) in continuum flux levels occur \citep{Martin04, Feast01}, and  the variations are not likely to be associated with a variation in dust extinction. \\
\indent The P-Cygni wind profile of a single massive star can be described by emission originating throughout the wind with absorption of the background emission in line-of-sight towards the observer. For a binary system the wind profiles are more complex due to excitation and motion of the companion star. In the case of \ec, lines such as in \ion{H}{1} and \ion {Fe}{2} are formed throughout the dominating primary wind. In the absence of a companion star, the \ion{He}{1} P-Cygni lines must form in a high-excitation zone close to the primary star. With a companion present, significant \ion{He}{1} emission  can arise in an ionized region near the bow-shock,  and hence at much larger radii. The structure is even more complex due to the highly eccentric binary orbit ($e$$\sim$0.9). The Coriolis effects lead to a highly distorted emission structure, especially near periastron passage. By contrast the \ion{He}{1} absorption arises from a confined  but asymmetric region in line-of-sight from the primary stellar source. Consequently, we have chosen to use the P-Cygni absorption component in our analysis. To determine the radial velocity variation of the absorption components we have fitted multiple Gaussians to the line profiles, as illustrated in Figure~\ref{f4_lab}. With this fitting procedure additional emission components are needed, which do not necessarily indicate spatially distinct emitting regions. The specific number of the narrow components used in the fit are chosen to mimic the general shape of the emission line and, as far as possible, to decouple the emission from the absorption.  The number of narrow components needed to fit each line profile changes with phase, but is consistent for all lines at a given phase. The accuracy of the measured radial velocity is phase dependent, due to a weak absorption or the possible presence of multiple components. Typical residuals to multi-Gaussian fits were less than two percent.

\begin{deluxetable*}{llcccc}
\tabletypesize{\scriptsize} 
\tablecaption{Spectral Lines  Used in This Analysis \label{t2}} 
\tablewidth{0pt} \tablehead{
\colhead{Spectrum} &
\colhead{\lm$_{vac}$} &
\colhead{Transition} &
\colhead{E$_\mathrm{low}$ } & 
\colhead{E$_\mathrm{up}$}  &
\colhead{$\log{gf}$} \\ 
\colhead{}&
\colhead{(\AA)} &
\colhead{}&
\colhead{(\cm)\tablenotemark{a}} &
\colhead{(\cm)\tablenotemark{a}} &
\colhead{}}
\startdata
\textup{He}\,\textsc{i}\tablenotemark{b} & 3188.67 & 2s $^3$S$-$4p $^3$P  & 159856 & 191217  & $-$1.16 \\
                                        & 4472.73 & 2p $^3$P $-$ 4d $^3$D & 169087 &191445   & \phs0.05 \\  
                                        & 5017.08 & 2s $^1$S$_0$ $-$ 3p $^1$P$_1$ & 166278 & 186209 & $-$0.82 \\
                                        & 5877.31 & 2p $^3$P $-$ 3d $^3$D  &169087 & 186102 & \phs0.74 \\     
                                        & 6680.00 & 2p $^1$P$_1$ $-$ 3d $^1$D$_2$  &171135 & 186105 & \phs0.33 \\  
                                        & 7067.20  & 2p $^3$P $-$ 3s $^3$S  &169087 & 183237 & $-$0.21 \\   
                                        & 7283.36 & 2p $^1$P$_1$ $-$ 3s $^1$S$_0$  &171135 & 184865 & $-$0.84 \\                                         
\textup{H}\,\textsc{i}\tablenotemark{c}   &  3723.00  &  2s $-$ 14p   &  82259  & 109119 & $-$1.98 \\     
                                        &  3735.43  &  2s $-$ 13p       &  82259  &  109030 & $-$1.87  \\       
                                        &  3751.22  &  2s $-$ 12p       &  82259  &  108917 & $-$1.76  \\
                                        &  3771.70  &  2s $-$ 11p       &  82259  &  108772 & $-$1.64 \\  
                                        &  3798.98  &  2s $-$ 10p       &  82259  &  108582 & $-$1.51  \\
                                        &  3836.47  &  2s $-$ 9p\phn &  82259  &  108324 & $-$1.36 \\
                                        &  4102.89  &  2s $-$ 6p\phn &  82259  &  106632 & $-$0.75 \\   
                                        &  4341.68  &  2s $-$ 5p\phn &  82259  &  105292 & $-$0.45 \\
                                        &  8865.22  &  3s $-$ 11p       &  97492  &  108772 & $-$1.05 \\
                                        &  9017.39  &  3s $-$ 10p       &  97492  &  108582 & $-$0.90 \\ 
\textup{Fe}\,\textsc{ii}\tablenotemark{d}  &  5019.84  &  4s$^2$ a$^6$S$_{5/2}$ $-$ 4p z$^6$P$_{5/2}$ & 23318  & 43239 & $-$1.35\\
                                                                        &  5170.47  &  4s$^2$ a$^6$S$_{5/2}$ $-$ 4p z$^6$P$_{7/2}$ & 23318  & 42658 & $-$1.25\\
\textup{N}\,\textsc{ii}\tablenotemark{b}  &  5756.19  &  2p$^2$ $^1$D$_2$ $-$ 2p$^2$ $^1$S$_0$&15316  & 32689 & $-$8.24 \\
\enddata
\tablecomments{Energy levels and wavelengths for \altion{H}{i} and the \altion{He}{i} triplets are weighted averages due to multiple unresolved transitions. The $\log{gf}$ is a sum over all transitions in the multiplet. All wavelengths are in vacuum.}
\tablenotetext{a} {1 \cm\  = 1.240$\times$10$^{-4}$ eV}
\tablenotetext{b} {Atomic data from \url{http://physics.nist.gov/PhysRefData/ASD/index.html}}
\tablenotetext{c} {Atomic data from \citet{Kurucz88}}
\tablenotetext{d} {$\log{gf}$ from \citet{R98}, all other atomic data from \citet{Kurucz88}}
\end{deluxetable*}

\section{Line Profile Variability \label{sect_lpv}}
The line profiles in the \ec\ spectrum exhibit a wide variety of shapes and variability. Below we discuss the general characteristics and observed changes.

\subsection{\ion{He}{1} Profiles \label{sect_hei_var}}
The \ion{He}{1} line profiles are complex and appear to be a composite of multiple features. The profiles change considerably between observations, especially in the phase interval 0.85$<$$\phi$$<$1.20 (Figures~\ref{f2_lab} and \ref{f3_lab}). On top of a broad weak emission are relatively narrow emission components that change rapidly with time. All spectral line characteristics (i.e. strength, shape, and radial velocity) vary systematically with phase. For each observation,  the \ion{He}{1}  profiles are shifted to the blue relative to the position of the \ion{H}{1} lines (compare Figures~\ref{f2_lab} and \ref{f3_lab} with Figures~\ref{f5_lab} and \ref{f6_lab}). While both the \ion{H}{1} and \ion{He}{1} emission weaken during the minimum, the velocity shift for the \ion{He}{1} emission is pronounced, and much larger than the corresponding shift observed in \ion{H}{1} (see Section~\ref{sec:h1}). We observe these shifts especially on the red side of the emission line where the profile is unperturbed by the blueshifted absorption. With phase, the \ion{He}{1} emission  increasingly shifts to the blue until just before the X-ray low-state begins.  Between phases 0.995 and 1.013, the intensity of the emission abruptly drops and the centroid of the emission line shifts to more positive velocities. After the minimum the emission is gradually more blueshifted. At some epochs the profile is nearly symmetric while at others the profile shows two or more narrow peaks. The limited sampling of this varying profile, and possibly spatially limited resolution, complicates the interpretation  of  the emission line radial velocities. Detailed modeling and possibly higher spatial/spectral resolution are required for a quantitative interpretation of the velocity variations.

\begin{figure}
   \centering
   \resizebox{\hsize}{!}{\includegraphics[angle=0]{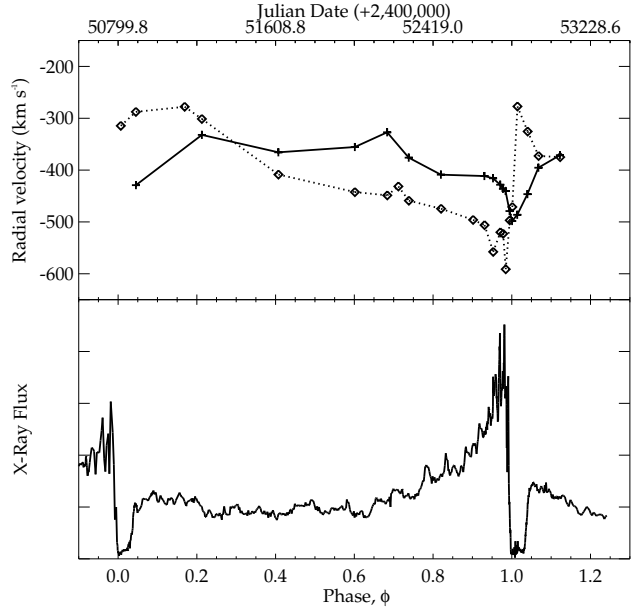}}
   \caption{Top: Radial velocities measured for the \altion{He}{i} absorption (diamond/dashed) and \altion{H}{i} (crosses/solid). Note the dramatic velocity shift for the \altion{He}{i}\ lines after periastron ($\phi$=1.0). Bottom: X-ray light-curve from \citep{Corcoran05} \label{f7_lab}.}
\end{figure}

\begin{figure}
\centering
\resizebox{\hsize}{!}{\includegraphics[angle=90]{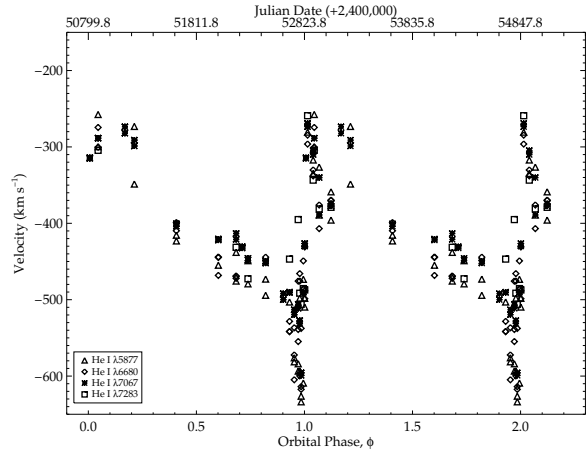}} 
\caption{Radial velocities for the \altion{He}{i} absorption. The typical measurement error is $\sim$20 \kms. Data points recorded between 0 and 1.2 are re-plotted one period later to show periodic variability. \label{f8_lab}}
\end{figure}

The \ion{He}{1} absorption components vary in velocity and strength and is traceable over \ec's entire 5.54 year spectroscopic period. The absorption velocity varies systematically (except between $\phi$=0.0$-$0.2) with a discontinuous jump of more than 200~\kms\ near periastron (Figure~\ref{f7_lab}).

\begin{figure}
\centering
\resizebox{\hsize}{!}{\includegraphics[angle=90]{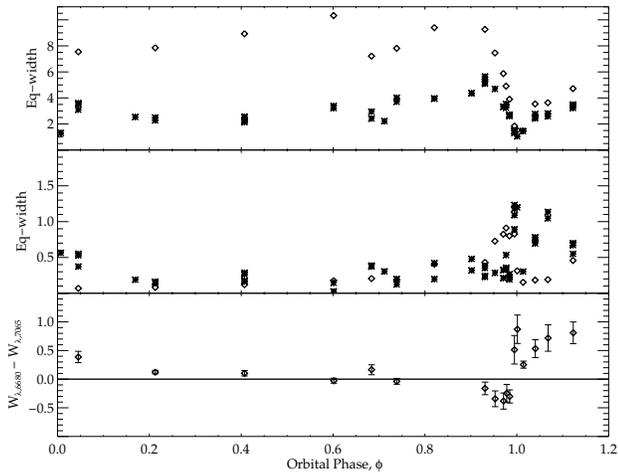}} 
\caption{Top: Emission equivalent widths for \altion{He}{i} \lm\lm6680, 7067. Middle: Absorption equivalent width for \altion{He}{i} \lm\lm6680, 7067. \altion{He}{i} \lm6680 shown by asterisks and \altion{He}{i} \lm7067 by diamonds. Bottom: Difference in absorption equivalent width as a function of phase. Error bars on the data points represent a 20\% uncertainty in  measurements for each of the two lines. The shift  over the spectroscopic cycle indicates a change in dominant excitation mechanism. During periastron an increase of equivalent width in \altion{He}{i}\ \lm7067 (triplet) relative to \altion{He}{i}\ \lm6680 (singlet) indicates more ionization/recombination relative to photoexcitation. \label{f9_lab} }
\end{figure}

\begin{figure}
   \centering
   \resizebox{\hsize}{!}{\includegraphics[angle=0]{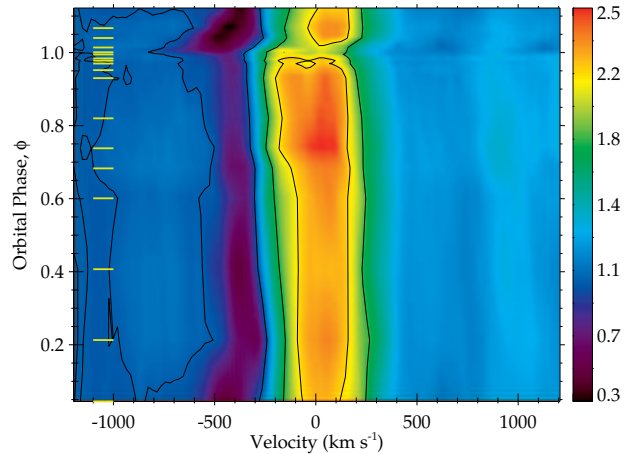}} 
   \caption{Surface plot for \altion{H}{i}\ $\lambda$3836. In contrast to \altion{H}{i}\ \lm4103, P-Cygni absorption is readily detectable at all phases. Like \altion{H}{i}\ \lm4103, and in contrast to the \altion{He}{i} lines, the \altion{H}{i}\ \lm3836 emission profile and radial velocity is relatively constant. Near $\phi$=1.0, the emission profile weakens significantly with an associated velocity shift.  The contours represent the flux level at $-$1200 \kms\ plus 0.70 and 0.85 of maximum line emission. \label{f10_lab}}
\end{figure}

In Figure~\ref{f8_lab} we show the measured absorption radial velocities for  \ion{He}{1} \lm\lm5877, 6680, 7067, 7283. The behavior previously discussed can be observed in all the lines. However, there is  scatter in the radial velocity measurements of different lines at the same phase. The absorption in \ion{He}{1}\ and \ion{H}{1}, apparent in Figures \ref{f3_lab} and \ref{f6_lab}, is strongest at different phases. The \ion{He}{1}\ absorption is strongest in the phase interval 0.8$<$$\phi$$<$1.0, and relatively weak at other phases. Conversely, the \ion{H}{1} P-Cygni absorption, particularly of the Balmer series members,  strengthens after $\phi$=0.98, is extremely strong at $\phi$=1.0, and  weakens after the minimum. \\
\indent \citet{Damineli00} investigated the variability in \ion{He}{1} \lm6680 emission using ground-based, seeing-limited spectra. They concluded that the changes with phase were caused by excitation effects rather than motion of either of the objects in the binary system. The velocity amplitude measured in \ion{He}{1} \lm6680 emission by \citet{Damineli00} is in qualitative agreement with our observations. However, the results from the \citet{Damineli00} analysis are based on variability in emission line strength while our results originate from absorption line radial velocities. The results may not be directly comparable but do show similar results regarding the amplitude and overall characteristic of the radial velocity curve. \\
\indent The relative strength  of the \ion{He}{1}\ triplet lines to the singlet lines  (see Figure~\ref{f9_lab}) provide information whether the lines are formed by photoexcitation or photoionization/recombination processes. An important question for understanding the origin of the \ion{He}{1}\ absorption is what determines its strength: is it primarily due to optical depth effects or does the fractional coverage of the continuum emitting source also play a role?  \\
\indent We comment on the \ion{He}{1} lines that can or cannot be studied in the \ec\  spectra covered by STIS:

\begin{enumerate}
\item Absorption from the metastable state 1s2s~$^3$P. All observable transitions from this term are  severely blended. \ion{He}{1} \lm3889 is, for example, blended with \ion{H}{1}~\lm3890. 

\item Absorption from 1s2s~$^1$S, such as \ion{He}{1} \lm5017. This line is blended with \ion{Fe}{2} \lm5020. However, the effect of the \ion{Fe}{2} line can be determined using \ion{Fe}{2} \lm5170 as \ion{Fe}{2} \lm\lm5020, 5170 have the same lower energy  state. Since the $gf$ value for \ion{Fe}{2} \lm5170 is similar to that of  \ion{Fe}{2} \lm5020, absence of significant absorption in \ion{Fe}{2} \lm5170  for any specific observation indicates that the absorption associated with \ion{Fe}{2} \lm5020 is due to the \ion{He}{1} line. From $\phi$=0.2$-$1.0,  the \ion{He}{1} contribution dominates the absorption, which is reasonable since the 1s2s~$^1$S  is a long-lived state. Late in this interval, the absorption depth exceeds 50\%, indicating that a significant fraction of the background emitting region is covered by  \ion{He}{1} gas responsible for the absorption. Unfortunately, from $\phi$=0.0$-$0.2 and 1.0$-$1.2, the \ion{Fe}{2} component becomes strong while, as shown by other \ion{He}{1} lines, the \ion{He}{1} \lm5017 absorption weakens. Hence, \ion{He}{1} \lm5017 cannot be used to trace the \ion{He}{1} absorption across the minimum. Absorption associated with \ion{He}{1} is seen on \ion{He}{1} \lm3966, which is a similar transition as \ion{He}{1} \lm5017,  but this line is severely  blended.

\item Absorption from 1s2p~$^3$P, including \ion{He}{1} \lm\lm4473, 5877, 7067, where \ion{He}{1} \lm5877 is the  strongest. The strength of the absorption lines are partially controlled by optical depth effects, and not dominated by effects related to partial coverage of the continuum source. 

\item Absorption from 1s2p~$^1$P, including \ion{He}{1} \lm\lm4923, 6680, 7283.  \ion{He}{1} \lm6680 is the least contaminated, but \ion{He}{1} \lm7283 is usable in this analysis.
\end{enumerate}

Analysis of the \ion{He}{1} absorption shows, that at some phases, the \ion{He}{1} in the He$^+$ zone absorbs much of the flux from the back-illuminating source, \eca. For example, on 2002 July 4 ($\phi$=0.820), the central intensity in \ion{He}{1} \lm5017 is less than 50\% of the adjacent continuum level. We note that the line absorption equivalent widths do not vary uniformly. There is a significant difference in variation between lines as shown in Figure~\ref{f9_lab} where the equivalent widths of \ion{He}{1} \lm6680 (singlet) and \ion{He}{1} \lm7067 (triplet) are compared. A significant change in their relative strengths occurs around $\phi$=0.9$-$1.1, indicating changes of the physical conditions of the absorbing regions. Figure~\ref{f9_lab} shows a greater equivalent width for \ion{He}{1} \lm7067 near periastron compared to \ion{He}{1} \lm6680, indicating ioniasation/recombination to be the major mechanism for populating the \ion{He}{1} energy states when \ecb\ is close to \eca. \\
\indent We created  radial velocity curves for the \ion{He}{1} absorption (Figure ~\ref{f8_lab}) in \ion{He}{1} \lm\lm5877, 6680, 7067, 7283, and a mean radial velocity curve (Figure \ref{f7_lab}). As with the peak absorptions, the most blueshifted of \ion{He}{1} and \ion{H}{1} occur at different phases. The greatest blueshift for \ion{He}{1} occurs leading up to the X-ray low-state ($\phi$=1.0).  In comparison, the greatest blue-shifts for \ion{H}{1} and \ion{Fe}{2} occur during and post minimum and are noticeably less than observed in the \ion{He}{1} lines.

\subsection{\ion{H}{1} Profiles \label{sec:h1}}
We use \ion{H}{1} \lm4103 (H$\delta$) to best illustrate spectral variations in hydrogen. This is a moderately intense, unblended line and is influenced by radiative transfer effects in a less complicated way than H$\alpha$. The H$\delta$ line, as presented in Figures~\ref{f5_lab} and \ref{f6_lab}, shows phase dependent variations in both emission and absorption. The H$\delta$ emission component varies substantially less than observed in \ion{He}{1} (see  Section~\ref{sect_hei_var}). Normalized to the continuum flux, the \ion{H}{1} lines do not vary greatly in intensity, although a change occurs close to periastron ($\phi$=1.0), where the peak emission in H$\delta$ drops by approximately 30\%. This behavior has previously been discussed by \citet{Davidson2005Ha}, where it was interpreted as evidence of a shell-ejection event. During the minimum, the H$\delta$ profile is flat topped ($\phi$=1.01 in Figure~\ref{f5_lab}), unlike profiles observed at other phases. One reason for invoking a shell-ejection is that a similar line profile was not observed during earlier spectroscopic minima. An alternative explanation is that \ec\ exhibits long term mass loss variations, and that these are superimposed on the phase related variations. The velocity variation of the emission is small ($<$25 \kms), except between phases 0.9 and 1.1, and is illustrated by the stable behavior of the red side of the emission profile. The velocity shift observed on the red portion of the \ion{H}{1} lines is, however, dependent on a good understanding of the continuum and the spectral line shape. The measurements of the position of the  absorption component yield a velocity amplitude of $\sim$60 \kms.  In an earlier study \citet{Damineli00}, using ground-based spectra, derived a radial velocity curve using  a large number of velocity measurements in \ion{H}{1} Pa$\gamma$ and Pa$\delta$. They derived a velocity amplitude of 50 \kms, centered at the systemic velocity. However, their orbital solution was, according to \citet{Davidson2000bin}, inconsistent with higher spatial resolution \hst\ spectra. A major difficulty in interpreting the velocity shifts in the hydrogen lines occurs because the P-Cygni profile is confused in ground-based data, by nebular scattered radiation and emission from the Weigelt blobs. \\
\indent The major part of the \ion{H}{1} emission is formed in the extended wind of \eca. Some of the variations and structure seen on top of the broad emission may arise from asymmetries in the  primary wind, or alternatively from changes in emission from the bow-shock associated with the wind-wind collision. Similarly, velocity variations could be due to the motion of \eca, or to changes in the wind structure. In both cases the observed amplitude of the velocity variations will be reduced since the \ion{H}{1} emitting region is extended well beyond \ecb's orbit. Consequently, it takes a finite time to replenish the gas. In the model of \cite{Hillier06}, roughly 50\% of the emission originates beyond $R$$\sim$1500\Rsun. A characteristic flow time ($R$/$v_A$) is thus 25 days, assuming $v_A$=500 \kms\ which is \eca's terminal wind velocity. Recombination time scales are much shorter. \\
\indent In contrast to the emission, the \ion{H}{1} P-Cygni absorption is highly variable. At some phases (e.g. just after $\phi$=1.0) the absorption is strong, while at other phases (0.50$<$$\phi$$<$0.95) it is much weaker, and in some lines difficult to detect. For H$\delta$ the most striking behavior of the P-Cygni absorption is the rapid strengthening near $\phi$=1.0. At $\phi$=0.984 (2003 June 1) no absorption can be seen in H$\delta$ while it is marginally detectable on H$\alpha$. By $\phi$=0.995 (2003 June 22), the absorption increased dramatically. P-Cygni absorption can be seen on high-n members of the Balmer series (Figure~\ref{f10_lab}) at all epochs observed with \hst\ STIS.

\begin{figure}
   \centering
   \resizebox{\hsize}{!}{\includegraphics[angle=0]{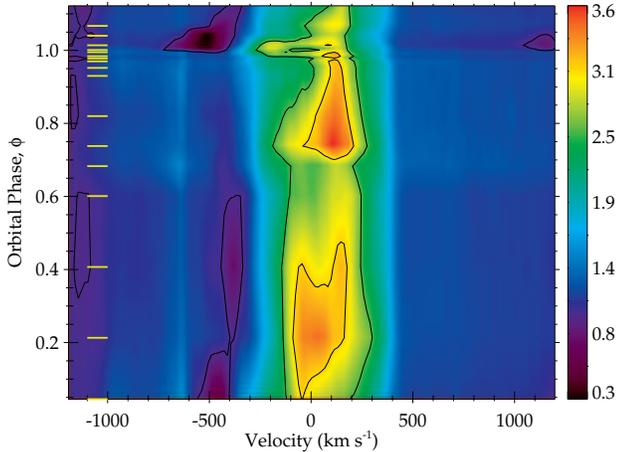}}
   \caption{Illustration of \altion{Fe}{ii}\ \lm5170 variability in intensity and velocity over the spectroscopic period. The absorption component, and emission on the blue side, are strongly variable with phase. Conversely, emission on the red side is relatively stable in strength, and in radial velocity.  The contours represent the flux level at $-$1200 \kms\ plus 0.7 and 0.85 of maximum line emission. \label{f11_lab}}
\end{figure}

\begin{figure}
   \centering
   \resizebox{\hsize}{!}{\includegraphics[angle=0]{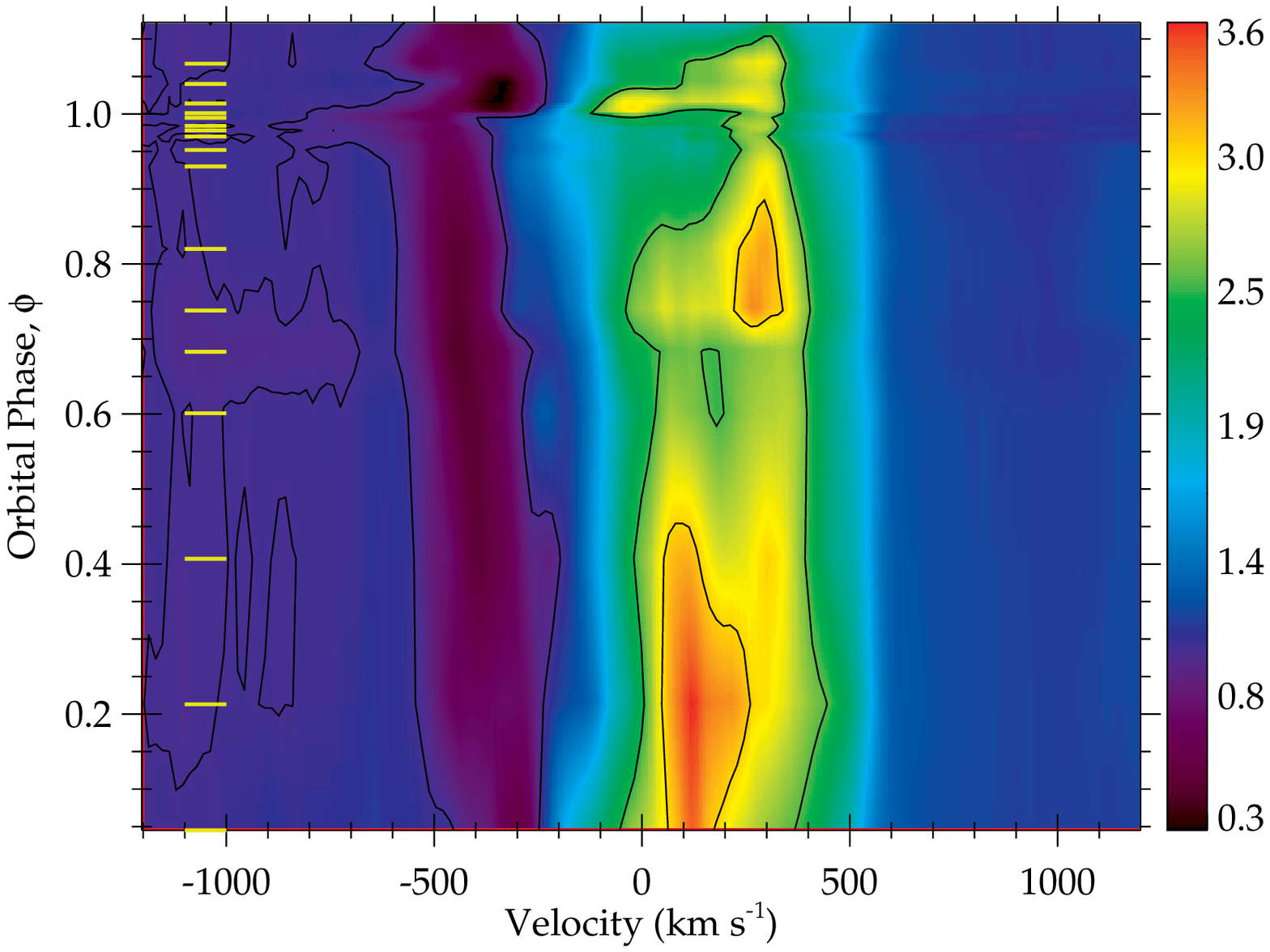}}
   \caption{Illustration of \altion{He}{i}~\lm5017 which is blended with the stronger \altion{Fe}{ii}\ \lm5020. The velocity scale is for \altion{He}{i}\ \lm5017.  \altion{Fe}{ii} \lm5020 has the same lower level as \altion{Fe}{ii}~\lm5170 (Figure~\ref{f11_lab}), and belongs to the same multiplet. Comparison shows that absorption in the \altion{He}{i} component dominates between $\phi$=0.2$-$1.0.  The contours represent the flux level at $-$1200 \kms\ plus 0.70 and 0.85 of maximum line emission. \label{f12_lab}}
\end{figure}

The \ion{H}{1} absorption shows radial velocity variations across the period (Figures~\ref{f6_lab},  \ref{f7_lab} and \ref{f10_lab}). During most of the period,  the \ion{H}{1}\ absorption velocity typically is between $-$330 to $-$400 \kms, but varies substantially around periastron. In the phase interval 0.95$<$$\phi$$<$1.20, the absorption radial velocity falls from approximately $-$400 to $-$500 \kms, before recovering to its pre-minimum level. 

\subsection{\ion{Fe}{2} Profiles}
The \ion{Fe}{2} profiles also show intensity and radial velocity variability. To illustrate the \ion{Fe}{2} variability we use \ion{Fe}{2} \lm5170 (see Figure~\ref{f11_lab}) which is one of the stronger \ion{Fe}{2} wind lines in the optical spectrum of \ec. According to models of luminous blue variables (LBVs) like \ec, the \ion{Fe}{2} emission is primarily produced by continuum fluorescence \citep[e.g.][]{Hillier98, Hillier01}. This line provides a useful comparison to the \ion{He}{1} \lm 5017 + \ion{Fe}{2} \lm5020 feature (Figure~\ref{f12_lab}). The third member of the multiplet, \ion{Fe}{2} \lm4925, shows a similar emission profile as \ion{Fe}{2} \lm5170 at all phases. \\
\indent Radial velocity variations are difficult to measure since the profile shows marked variability on the blue side. While some of this variability appears to be directly related to the P-Cygni absorption, much of it can be associated with change in the emission on the blue side. Conversely, the red side of the profile, with the exception of the spectrum recorded at phase 0.21 (1999 February 21) shows minimal variability, with the velocity changes typically being less than 20 \kms. Similar to the \ion{H}{1}\ line profiles, the \ion{Fe}{2} absorption components are weak throughout most of the 2024 day cycle. They begin to strengthen around $\phi$=0.98, are strong just after periastron ($\phi$=1.0), but have virtually disappeared at $\phi$=1.2. \\
\indent Given the strength of the \ion{Fe}{2} emission in the \ec\ spectrum, the weak P-Cygni absorption in all profiles of lines in the multiplet,  including \ion{Fe}{2} \lm\lm5020, 5170,  throughout the broad maximum is difficult to explain with a spherically symmetric model. At times, the \ion{Fe}{2} emission is double-peaked, while at other occasions the blue side of the profile is severely weakened relative to the red side.\\ 
\indent The spectral behavior observed in \ion{Fe}{2} can indicate distinct emitting regions, one of which is highly variable. Alternatively, the bulk of the \ion{Fe}{2}\ emission could come from the primary stellar wind. Variations would then be explained by significant perturbations to the ionization structure of the wind region that affects the blueshifted emission. Such a scenario is consistent with a hot companion star ionizing the wind on the observer's side of \eca. 

\subsection{[\ion{N}{2}] Profiles}
\begin{figure}
   \centering
   \resizebox{!}{\hsize}{\includegraphics[angle=0]{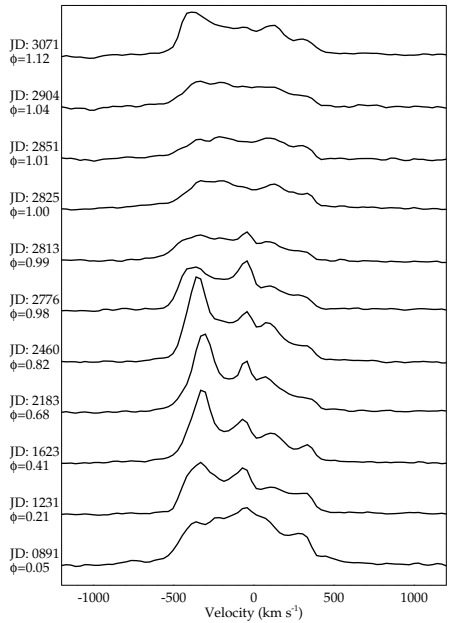}} 
   \caption{[\altion{N}{ii}] \lm5754 variations over the spectroscopic period. The line profile is asymmetric and can be described with four narrow components which mutually vary in intensity but not in velocity  with phase. The line is centered near at the $-$8 \kms\ system velocity. There may be a broad underlying component associated with the stellar wind. \label{f13_lab}}
\end{figure}

\begin{figure}
   \centering
   \resizebox{\hsize}{!}{\includegraphics[angle=0]{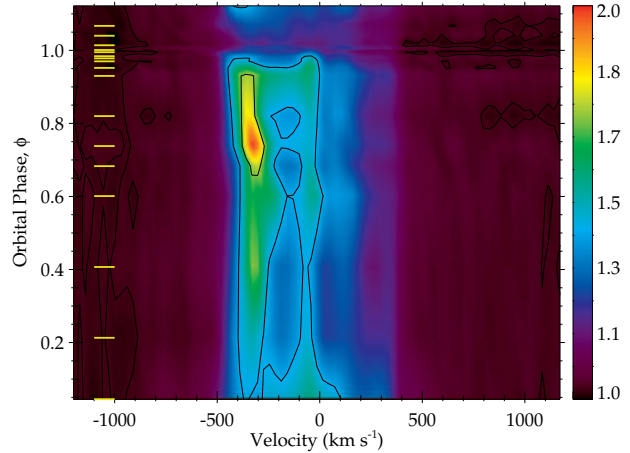}}
   \caption{Illustration of the [\altion{N}{ii}] \lm5756 variability in intensity and velocity over the spectroscopic period. The contours represent the flux level at $-$1200 \kms\ plus 0.70 and 0.85 of maximum line emission. \label{nii_surf}}
\end{figure}

The [\ion{N}{2}] \lm5756 line shows a complex structure with multiple narrow emission peaks spanning velocities of  $-$500 to +500 \kms  (Figures~\ref{f13_lab} \& \ref{nii_surf}). This line profile is significantly broader than the \ion{He}{1} emission lines and is centered about the $-$8 \kms\ system velocity. Near $\phi$=1.0 the emission is weak, with the observed line representing intrinsic [\ion{N}{2}] emission from the primary wind. At other phases the emission is stronger, more structured with the emission weighted towards the blue. Such emission can arise due to additional ionization of the primary wind by \ecb, or line formation in the bow-shock. At phase 0.21 and 0.68 the [\ion{N}{2}] profiles differ substantially from those in \ion{He}{1}. The emission is more heavily weighted to the blue side. The [\ion{N}{2}] emission components show little or no velocity variation; only the intensity of individual components changes with orbital phase. The lack of velocity variation indicates that the [\ion{N}{2}] emission is formed beyond the inner, denser portions of the primary wind or the wind-wind collision region. The [\ion{N}{2}] emission, as with numerous [\ion{Fe}{2}] lines and other forbidden line species,  originates in lower density, outer wind regions. Using the model of \citet{Hillier01} the extended [\ion{N}{2}] emitting region coincides with, and extends well beyond, the binary orbit (semi-major axis $\sim$15 AU). While the critical density of the $^1$D state is approximately $8.6\times10^4$\,cm$^{-3}$ at $10^4$\,K \citep{Osterbrock2006}, most of the emission may come from larger particle densities if hydrogen has already recombined.

\section{Single Star Shell-Ejection vs. Binary driven Variability}
A proper interpretation of the line profile variability depends on the excitation mechanisms in the line formation region.  Since \ec\ is considered to be an LBV, a natural interpretation is that the observed emission and absorption line variations are evidence for periodic shell-ejections. Shell-ejections were invoked, for example, by \citet{Zan84_shell} to explain the observed variation of high-excitation features seen in ground-based spectra. The discovery of the 2024 day period suggested that the variations could be explained by orbital modulation in a binary system \citep[e.g.][]{Damineli97, Damineli98}. In this model, flux from \ecb\ ionizes more completely a portion of  the outer wind on the side of \ecb\ and weakens the hydrogen absorption. For this model to work, our line-of-sight must intersect the ionized zone\footnote{By ionized we mean H/H$^+$$\sim$10$^{-6}$. In this case the Balmer P-Cygni absorption tends to be weak. When H/H$^+$$\sim$10$^{-3}$ or larger, the absorption is strong.}. At periastron the flux from \ecb\ is absorbed/occulted by \eca\ and its inner wind, and the \ion{H}{1} absorption lines are stronger.\\ 
\indent A single star shell-ejection model is still invoked by some authors to explain the observed line and continuum variations \citep[e.g.][]{Davidson2005Ha}. \citet{Smith03b} invoked shell-ejection to explain the observed variation of the \ion{H}{1}\ P-Cygni profiles. In their model, the central source has a fast polar wind which is denser than the equatorial wind. At most phases the equatorial wind is highly ionized and the \ion{H}{1} absorption is weak. However, during the spectroscopic event a shell is ejected and the wind becomes symmetric with enhanced \ion{H}{1} absorption.  The systematic variation of the \ion{H}{1} absorption profiles seen in radiation reflected by dust in the Homunculus and the symmetry about the polar axis, support this model.\\
\indent Both models can potentially explain the observed variations of the Weigelt blob spectra. In the single star shell-ejection model, radiation from the central source preferentially escapes in the equatorial regions, ionizing the Weigelt blobs. During the shell-ejection, this radiation is blocked, and hence the ionization of the Weigelt blobs decreases.  In the binary model, \ecb\ supplies the harder photons necessary to produce \ion{He}{1}, \ion{Fe}{3} etc. When \ecb\ approaches periastron, \eca's extended wind and perhaps \eca\ itself, shields the blobs from the ionizing radiation. \\
\indent Can we distinguish between the single star shell-ejection model and the binary model? We interpret the present spectroscopic data as strong support for binarity over a single star shell-ejection:

\begin{enumerate}

\item The \ion{He}{1} emission lines are consistently blueshifted throughout most of the 5.54 year period. This is not explainable with an axial-symmetric shell-ejection which would give equal amounts of blueshifted and redshifted emission. It is naturally explained by the binary model. In the binary model the \ion{He}{1} emission is produced by ionization by the far-UV flux from \ecb\ followed by recombination. The orbit is highly elliptical and at apastron (and for most of the orbit) \ecb\ is in front of \eca, and ionizes the portion of \eca's wind flowing towards the observer.  Therefore, for most of the orbit, the \ion{He}{1} is formed in the blueshifted part of \eca's wind.

\item  The double peak profile seen in the \ion{He}{1} emission (see Figure~\ref{f2_lab}), and the velocity variations of these peaks, are qualitatively consistent with emission arising from spatially separated regions near a colliding wind interface \citep{Luhrs97}.  The shape and behavior of the \ion{He}{1} emission lines in the \ec\ spectrum are similar to the line variations observed in other colliding wind binaries. The \ion{C}{3} \lm5698 line is observed to show dramatic line profile variation in the spectra of WR~42, WR~48 \citep{Hill00, Hill02} and WR~79 \citep{Luhrs97, Hill00, Hill02}. \ion{C}{3} \lm5698 is not present in the \ec\ spectrum.   

\item As noted earlier, a striking characteristic of the \ion{He}{1} emission is the dramatic velocity shift across periastron passage.  This is similar to radial velocity variations of stars in highly eccentric binary systems, and suggests the emission follows the orbit of one of the stars. In general it is difficult to see how a single star shell-ejection model can explain the complex behavior of the observed \ion{He}{1}\  absorption profiles (see Section~\ref{sec:rvsoln}). Potentially, the velocity jump near $\phi$=1.0 can be caused by a shell-ejection, with the absence of the higher velocity absorption from the previous shell disappearing because of recombination. In a single star shell-ejection model, radial velocity variations in absorption would occur since the ejected material accelerates to more negative velocities as it moves away from the star. However, in this model it is hard to explain why the most rapid decrease in velocity occurs close to 2003.5 minimum rather than just after the shell-ejection, which would be necessary to provide significant absorption for the entire 5.54 year cycle. Furthermore, it is not easy to explain why the \ion{He}{1} absorption increases in strength from $\phi$=0.80 in a single star shell-ejection model, since this is well before a shell is ejected. 

\item  While the increase in strength of the \ion{H}{1} absorption can be explained by a single star shell-ejection model, it is difficult to understand why the \ion{He}{1} absorption has its greatest blue-shift at $\phi$=1.0. In most single star shell-ejection models, the shell produces low-velocity absorption just after ejection. The observed velocity change occurs across $\phi=1.0$ and thereafter requires that the shell be ejected at a very high velocity ($\sim$$-$600 \kms) and decelerate just after $\phi$=1.0 (to $\sim$$-$300 \kms) before re-accelerating.  There is no obvious physical explanation for such behavior, other than the radial velocity variations are caused by orbital motion.

\item Another concern is that the \ion{H}{1} emission weakens near the time of the shell-ejection, before recovering to pre-2003.5 levels, but ejection of a shell should produce more \ion{H}{1} emission.  In the binary model the weakening can be associated with the disruption of the primary wind by the companion star close to periastron.

\item While high-n members of the Balmer lines show significant variations, P-Cygni absorption is present for these lines at all phases, implying that \ecb\ has a greater influence on the outer wind where, for example, H$\alpha$ is formed.

\item The variability in \ion{Fe}{2} can be naturally explained by a binary model. In the model of \citet{Hillier06}, \ion{Fe}{2} emission originates at large radii, extending beyond the binary orbit (30 AU). The companion's far-UV emission will enhance the ionization of nearby outflowing material, weakening both the emission and absorption. Because the orbit is highly elliptical, for most of the orbit \ecb\ is on the observer's side of \eca. On the other hand, a single star shell-ejection model can explain some of the observed variations if the ejected shell obstructs UV photons from \eca, lowering the degree of ionization and enhancing the strength of the \ion{Fe}{2} P-Cygni absorption. However, an axial-symmetric shell-ejection usually causes emission variations on both the blue and red sides of the line profile. Only strong variability on the blue side is observed. 

\end{enumerate}

There are still uncertainties and some of the individual points raised above are debatable. However,  the consistency between the different line profile diagnostics, the qualitative variation of the observed X-ray fluxes, the variation of the line intensities in the Weigelt blobs, and the location of the Weigelt blobs on the same side of \eca\ as \ecb\ at apastron, strongly support the binary scenario. Furthermore, the data suggest that the observed variations are a direct consequence of the variability of the UV radiation and wind of \ecb.

  \begin{figure}
   \centering
  \resizebox{\hsize}{!}{\includegraphics[angle=90]{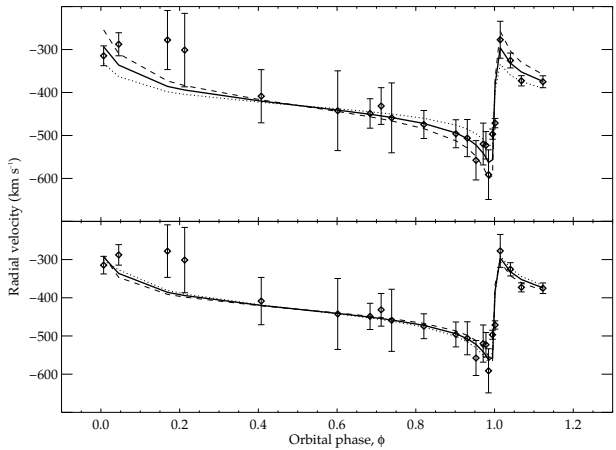}} 
   \caption{Radial velocity as a function of phase derived from the \altion{He}{i} absorption.  The solid curve shows a fit to the data with the parameter given in Table~\ref{t3} ($K$=140 \kms, $e$=0.90). Upper panel:  Variation in the velocity amplitude, $K$=100 \kms\ (dotted) and $K$=180 \kms\ (dashed).  Lower panel: Variation of the eccentricity,  $e$=0.88 (dotted) and $e$=0.92 (dashed). The error bars for the measurements represent an estimated uncertainty of the fit and is proportional to the 
inverse of measured equivalent width. \label{f14_lab}} \end{figure}

\section{Interpretation of the Velocity Variability\label{sect_ivv}} 
We have established that line profile variations observed over \ec's spectroscopic cycle are consistent with a binary model. However, a quantitative analysis of the observed variations is difficult because the lower excitation lines are formed in an extended part of the system. An exception is the \ion{He}{1} absorption, that presumably is formed in a localized region close to the wind-wind interface. The radial velocity variations in the \ion{He}{1} absorption appear to be reminiscent of radial velocity variations from a star in a highly eccentric orbit: a gradual change in radial velocity through most of the cycle, with a rapid shift near periastron passage. We therefore explore what these velocity variations tell us about \eca, \ecb, and the interacting gas. \\
\indent We show how the \ion{He}{1} absorption line variations  can be described by a standard eccentric binary radial velocity curve, and demonstrate that the derived value of the eccentricity is consistent with the value derived from the X-ray light-curve. To interpret the absorption components and the meaning of the velocity curve we present a detailed discussion of the colliding wind model and discuss the origin of the \ion{He}{1}\ emission and absorption components. 

\subsection{The \ion{He}{1}\ Radial velocity curve}
Line profiles recorded at several phases indicate that the \ion{He}{1} absorption is a composite of multiple components.  However, only the strongest, least blueshifted feature, is traceable over the entire spectroscopic cycle. We created a radial velocity curve for  the \ion{He}{1} absorption, based on this component. A second high velocity absorption component appears near the spectroscopic minimum (see Figure~\ref{f4_lab}) and disappears during the recovery phase. \citet{Stahl05} noted a higher velocity component that appeared near the time of the X-ray low-state, and claimed it was a result of a shell-ejection. Over the brief period in which it was observable it followed a velocity pattern similar to the less blueshifted component, suggesting formation in adjacent regions. We have compared the velocity variation with the radial velocity curves for \ion{He}{1} and \ion{H}{1} emission derived by \citet{Damineli00} and  to the X-ray light-curve \citep{Corcoran05}.  Measured absorption velocities for  the \ion{He}{1} and \ion{H}{1} are presented in Figure~\ref{f7_lab} along with a plot of the \rxte\  X-ray flux (2$-$10 keV). The maximum velocity of the \ion{He}{1} absorption is about $-$600 \kms, which decreased to about $-$300 \kms\ on 2003 June 22,  just before the spectroscopic minimum. The striking difference between the velocity variation for the \ion{He}{1} and \ion{H}{1} wind lines is the drop in velocity after periastron ($\phi$=0.05) which follows the X-ray drop seen by \chandra\ \citep{Corcoran06}. This discontinuity  may be a result of absorption in the gas between the primary star and the bow-shock, while the \ion{H}{1}  absorption originates from a volume extended beyond \ecb's orbit. The variation of the \ion{He}{1} absorption component is similar to that of a spectroscopic binary.  Figure~\ref{f14_lab} shows the observed \ion{He}{1} absorption velocities, along with sample velocity curves for a range of eccentricities and velocity amplitudes. Note that the spectral lines in this analysis have  absorption components which are blended with their emission components,  making the line profile fitting difficult and decreasing the measurement  accuracy.  The radial velocity curves for \ion{He}{1}, plotted in Figure~\ref{f14_lab}, were calculated using the standard formalism \citep{aitken}: 

\begin{equation}
\frac{dz}{dt}=K(e\cos{\omega}+\cos{(\nu+\omega}))+\gamma
\end{equation}

\noindent where $dz/dt$ is the radial velocity along the line-of-sight, $e$ is the eccentricity, $\omega$ the longitude of periastron, $\nu$ the true anomaly\footnote{The relation between the true anomaly, $v$, and the eccentric  anomaly, $E$, is given by $\tan{(\frac{v}{2})}=\sqrt{\frac{1+e}{1-e}} \tan{(\frac{E}{2})}$. $E$ is related to the phase as 2$\pi\phi=E-e\sin{(E)}$.} and $\gamma$ the average velocity.  $K$ is: 

\begin{equation}
K=\frac{2\pi a\sin{i}}{P\sqrt{1-e^2}}
\end{equation}

\noindent where $i$ the inclination, $a$ the major axis of the orbit and $P$ is the spectroscopic period with an assumed beginning at the X-ray low-state coinciding with periastron passage. All other parameters are initially free, although we constrain $e$$>$0.8 based on the solution of the X-ray light-curve \citep{Corcoran01}. In Figure~{\ref{f14_lab} solutions obtained by varying $e$ and $K$ are plotted. The adopted parameters are given in Table~\ref{t3}. 

\begin{figure}
 \centering
  \resizebox{\hsize}{!}{ \includegraphics{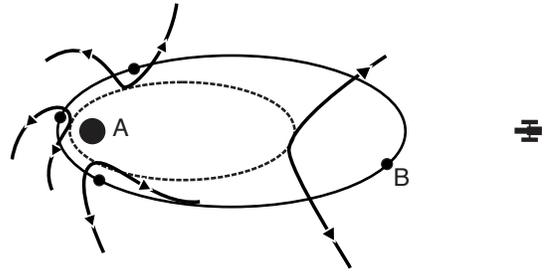}}
   \caption{A cartoon representing the standard view of the colliding wind system in \ec, in which the shock-cone is wrapped around \ecb\ (B). The more massive star \eca\ (A) has the slower but denser wind. The arrows show the flow of the gas along the bow-shock. The position of \hst\ is appropriate for $\omega$=270$^\circ$ and the system is likely tilted about 41$^\circ$ into the sky-plane.}
   \label{f15_lab}
\end{figure}

\begin{deluxetable}{cc}
\tabletypesize{\scriptsize} 
\tablecaption{Radial Velocity Curve Solution \label{t3}} 
\tablewidth{6cm} 
\tablehead{
\colhead{Parameter } &
\colhead{Value}}
\startdata
P & 2024.0 days\tablenotemark{1} \\
$i$ & $41^{\circ}$\tablenotemark{2} \\
$\gamma$ & $-$430 \kms \\
$K$ & 140 \kms \\
$e$ & 0.9 \\
$\omega$ & \phn270$^{\circ}$
\enddata
\tablenotetext{1}{\citet{Corcoran05}}
\tablenotetext{2}{\citet{Davidson01}}
\label{tab:params}
\end{deluxetable}

\subsection{Orbital Dynamics}
In most published binary models of the \ec\ system, \eca\ is defined as the primary star and is dominating both the mass and luminosity of the system. \ecb, is usually assumed to be the hotter, less massive star with a weaker, although much faster wind. Support for such a scenario comes from the analysis of the X-ray \citep{Corcoran06}, the far-UV \citep{Iping05} and UV/optical \citep{Hillier01, Hillier06} spectra in addition to the modeling of the Weigelt blobs \citep{Verner05a}. Though there is no direct evidence regarding the orbital inclination or the relation of the orbit to the geometry of the Homunculus,  the usual assumption is that the inclination of the orbit is the same as for the symmetry axis of the Homunculus, and that the orbital plane lies in the intervening disk between the lobes of the Homunculus. Most models \citep{Corcoran01, Smith04b} assume that the longitude  of periastron is such that, prior to periastron passage, \ecb\ is moving away from us and \eca\ is moving towards us, as shown in Figure~\ref{f15_lab}. \citet{Falceta05} suggested that \ecb\ is moving towards us just prior to periastron passage. It is more certain that the system has a large eccentricity and that, because of the short duration of the pan-chromatic variations, periastron passage occurs in the time interval of the X-ray/ionization low-state. \\
\indent The \ion{He}{1} P-Cygni absorption components are the best localized features in the system since they require high energy photons and/or collisions to be produced and they arise in out-flowing material distributed along the line-of-sight to the more luminous star. Nevertheless, the interpretation of these features is not straightforward since it is not well understood where the absorbing material is located. The absorber is either associated with the wind of \eca, with the colliding wind interface, the wind of \ecb, or a combination of these. Similarly, the \ion{He}{1} emission can either arise in the bow-shock or in the portion of the primary wind ionized by \ecb. Downstream absorption in the bow-show has previously been observed in other binary systems such as V444 Cyg \citep{Shore88}.  However, the thickness of the bow-shock in the \ec\ system and the projection in line-of-sight provide little absorbing material between \eca\ and the observer throughout the orbital period. The ionizing photons of \ecb\ penetrate deeply into \eca's wind creating a substantial He$^+$ zone. In the following sections we describe the possible location of the absorbing system, and discuss implications on the derivation of the system parameters. 

\subsubsection{Velocity variations caused by orbital motion\label{sec:rvsoln}}
The velocity curve shown in Figure~\ref{f14_lab} is very similar to a radial velocity curve for an object in a highly eccentric orbit. If Figure~\ref{f14_lab} describes the orbital motion of one of the two objects in the \ec\ system then the blueshifted \ion{He}{1} absorption over most of the spectroscopic period and the geometry of the system (including the position and variability of the Weigelt blobs), imply that we are tracing absorption in the dense primary wind. The terminal velocity of the absorption profile is consistent with the observed terminal velocity of \eca\ and much lesser than the terminal velocity of \ecb\ ($\sim$3000 \kms) \citep{Pittard03}. Wind profiles with a terminal velocity as large as 3000 \kms\ are not observed in the \ec\ spectrum. \\ 
\indent The derived radial velocity curve (Figure~\ref{f14_lab}) confirms the large eccentricity derived from the X-ray light-curve \citep{Corcoran01}, but yields a surprisingly large $K$-value, $K=140$ \kms. This value is inconsistent with the radial velocity amplitude measured in \ion{H}{1}, whose origin is associated with \eca. To determine a mass of both stars in the binary system, radial velocity curves for both objects are needed. However, if only one of the objects is traced then a mass function describing the relation between the two objects can be created. If the large \ion{He}{1} velocity shifts during periastron are solely caused by the orbital motion of one of the stars, then the mass  function is:

\begin{eqnarray*}
f(m_{1},m_{2}) & = & \frac{K_1^{3}P (1-e^{2})^{3/2}}{2\pi G} =  \frac{m_2^3\sin^3{i}}{(m_1+m_2)^2}\\
                           & = & \frac{m_1\sin^3{i}}{q(q+1)^2} = 50 M_{\odot}
\end{eqnarray*} 

\noindent using the values from Table~\ref{t3}, and $q$=$m_1/m_2$ with the subscripts denoting the observed star (1) and the unseen star (2). With the exception of the possible detection of \ecb\ in the \fuse\ spectrum \citep{Iping05}, there is little evidence for spectral features from the companion star. We therefore, initially, associate the \ion{He}{1} absorption with the primary star. If the observed velocities represent the motion of \eca, a mass for \ecb\ for chosen values of \eca's mass can be derived. This analysis is consistent with a mass of \ecb\ of $210M_{\odot}$ for a $\sim$20$M_{\odot}$ mass of \eca, i.e. \eca\ is the lower mass object in the system with an extreme mass ratio, $q$$\sim$0.1. This is the solution with the lowest total mass of the system, but is only one of many possible solutions with the set of parameters given in Table~\ref{t3}. The orbital plane is assumed to be in the same plane as the intervening disk between the Homunculus, $i$=41$^\circ$, but this value is poorly determined. If using a limiting value of the inclination closer to 90$^\circ$ a different set of solutions with a higher mass ratio (0.4$-$0.6) is obtained, however, \ecb\ still is the more massive object in the system. This is different than the conventional view of the system in which \eca\ has a mass of $\sim$100$M_{\odot}$ and is the more massive object. \\ 
\indent If the conventional view, as we believe, is correct, it requires that the mass function we derive above is a gross overestimate of the true mass function. This could be due to underestimation of the eccentricity or inclination, or an overestimate of the velocity amplitude, $K$, or a combination thereof. 
Note, that the argument that the more luminous star is the more massive one, assumes the two objects to be unevolved. If this is not the case and for example \eca\ is on the helium burning main sequence, then \ecb\ may be the more massive object in the system. \\
\indent In the next section we describe how ionization effects, caused by \ecb,  may influence the derived value of $K$.

\subsubsection{The Influence of Ionization on the Velocity Amplitude\label{rv_i}}
If the \ion{He}{1} absorption is in states populated by recombination in a region of \eca's wind, which is ionized by \ecb\ and localized towards \eca, then the observed absorption line velocities are a combination of \eca's wind velocity in the absorbing region in addition to the velocity due to \eca's  orbital motion.  Variations in the observed line velocities can either be produced by changes in the orbital velocity of the absorbing gas, in the characteristic outflow velocity of the gas, or a combination thereof.

\begin{figure*}
\centering
\includegraphics[angle=0,scale=0.55]{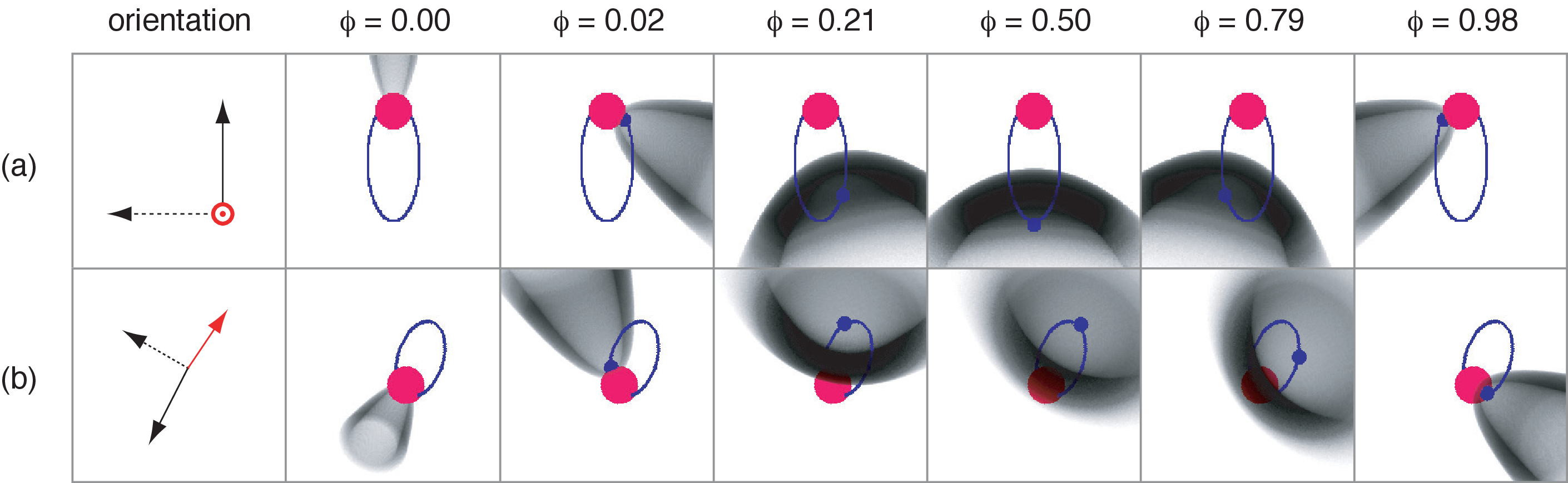}    
\caption{ Visualization of the He$^+$ zone located in the \ec\ system. \eca, the larger red stellar component, is shown with a $\sim$9 AU diameter disk, consistent with the apparent 2 $\mu$m continuum diameter measured by \cite{Weigelt06}. \ecb, the source of the UV radiation, is the smaller blue disk. The gray structure represents the He$^+$ zone, the source of the \altion{He}{i} emission and absorption. Top row (a) views the system with the plane lying in the paper: the Z-axis (red) points out of the paper, the X-axis (dashed arrow) and the Y-axis lie in the paper.  The semi-major axis of \ecb\ is in the Y direction and the semi-minor axis in the X direction. Bottom row (b) views the system with the Z-axis (red) pointed along the axis of the bipolar Homunculus  with Z  at $-$41$^\circ$, rotated about $-$45$^\circ$ in position angle from north, consistent with the geometry derived by \cite{Davidson01}.  Given the locations of the Weigelt blobs plus the orientation of the radio continuum monitored by \cite{Duncan03}, the orbit is rotated into the disk plane such that the semi-major axis (Y direction) projects almost on top of the Z-axis. NOTE: this visualization does not take into account distortion due to the Coriolis force which is considerable near periastron ($\phi$=0.00), but small near apastron ($\phi$=0.50).
\label{f16_lab}}
\end{figure*}

The location of the ionized region in \eca's wind is difficult to determine accurately since it depends on the density profile through its distorted wind, combined with the size of the wind-wind collision shock-front which allows more of \ecb's ionizing flux to penetrate into \eca's wind.  We illustrate an undistorted He$^+$ zone for different phases throughout the 5.54 year period in Figure~\ref{f16_lab}. We note that even the two-dimensional model of the colliding wind region becomes highly distorted by the Coriolis force near periastron (J. Pittard, 2006, private communication). Distortion of the wind-wind interface will greatly affect the derives radial velocity measurements, and the derived $K$ value. Building a detailed three-dimensional model is beyond the scope of this paper. The model shown in Figure \ref{f16_lab} simply approximates the wind-wind interface and the He$^+$ boundary as two undistorted parabolical surfaces centered on the line between the two stars. \\
\indent Figure~\ref{f16_lab} (top row) shows the system as viewed looking down on the orbital plane. We see that, other than near periastron ($\phi=0$), the He$^+$ structure is predominantly offset from \eca\ towards \ecb. In this orientation, the velocity seen in emission will be centered at the system velocity, but the observer would see a range in velocities dependent upon the angular curvature of the bow-shock in three dimensions.  The major part of the \ion{H}{1}~Lyman radiation escapes down the direction of the major axis and is directed in a small cone along the orbital plane toward the Weigelt blobs. \\
\indent The bottom row in Figure~\ref{f16_lab} shows the binary system viewed by the observer, assuming an inclination, $i$=$-$41$^\circ$, which is appropriate if the orbital plan is in the disk plane between the bipolar lobes of the Homunculus. The orientation depicted in the bottom row of Figure~\ref{f16_lab} is such that for most of the period, a significant portion of the He$^+$ zone overlaps the line-of-sight towards \eca, but at different radial velocities if the wind is accelerating outwards from \eca.

\begin{figure}[htbp] 
   \centering
   \includegraphics[width=3in]{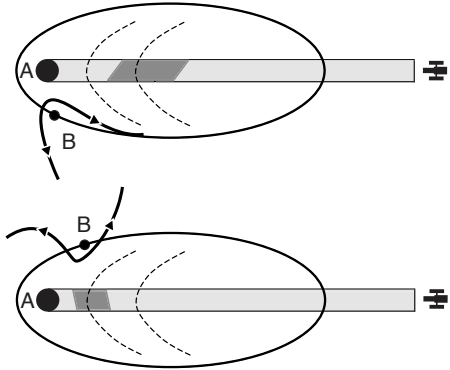} 
   \caption{ A qualitative illustration showing the orbit of \ecb\ in the reference frame of \eca\ and how the ionization of \eca's wind by \ecb\ can produce absorption line velocity variations. The dashed lines represent wind isovelocity contours. The dark grey area represents the section of \eca's wind which is ionized by \ecb's flux. Prior to periastron passage (\textit{top}), \ecb\ primarily ionizes high-velocity wind material far from \eca.  After periastron passage (\textit{bottom}) \ecb\ mostly ionizes low velocity wind material due to a buildup of cool wind material between the observer and \ecb. This is a qualitative representation and not drawn to scale.}
   \label{{f17_lab}}
\end{figure}

Figure~\ref{{f17_lab}} qualitatively shows the effects of ionization on the material in front of the disk of \eca. The rectangular region represents the material towards \eca, and the dark grey region represents the He+ zone and scales with the \ion{He}{1} column density, proportional to N(He$^{+}$)$\times$N(e$^{-}$). Prior to periastron passage, the companion star mostly ionizes high-velocity wind material far from the primary in the direction of the shock-cone, which points away from \eca. After periastron passage, the shock-cone is tilted towards \eca, so that \ecb\ mostly ionizes low velocity material in \eca's inner wind. X-ray spectra also show a buildup of cool, absorbing material between the colliding winds and the observer for a brief time after periastron passage \citep{sax02, Corcoran05}, which blocks ionizing radiation from the companion star from reaching the outer, high velocity wind material. Thus, after periastron passage  the observed velocity of the absorption system is expected to shift to lower velocities.  This means that ionization effects can produce larger velocity variations than would be observed from orbital motion alone, and therefore the velocity amplitude derived in Section~\ref{sec:rvsoln} does not represent the actual velocity variation due to the orbital motion of \eca. If the ionized structure was known, the \ion{He}{1} radial velocity curve could be corrected for the influence of ionization and the true orbital motion of \eca\ could be derived. However, this requires detailed modeling of the dynamics of the wind of \eca\ and \ecb.  

 \section {Discussion}
Evidence from numerous observations suggest that \eca\ is orbited by a hot companion star in a highly eccentric orbit. The velocity curve derived from the \ion{He}{1} P-Cygni absorption, and to a lesser extent, the variation of \ion{H}{1} absorption show periodic behavior very similar to the radial velocity of a star in a highly eccentric orbit with a longitude of periastron $\omega$$\sim$270$^{\circ}$, which is consistent with the orbital parameters derived from analysis of the \rxte\ X-ray light-curve \citep{Corcoran01}, and also confirm the variation seen in ground-based observation of the Pa$\gamma$ and Pa$\delta$ emission lines \citep{Damineli00}. Significant deviations from $\omega$ by more than about 10$^{\circ}$ could not describe the observed radial velocity curve.  Our solution requires a much larger eccentricity than the solution given by \citet{Damineli00}, $e$=0.9 instead of $e$=0.65$-$0.85. This larger eccentricity helps resolve the discrepancy between the ground-based Paschen line velocities and an earlier report on Paschen line velocities in two observations with \hst\ STIS \citep{Davidson2000bin}. In any event the more complete STIS dataset analyzed in this paper clearly show radial velocity variations of the type expected from orbital motion. Our analysis of the radial velocity curve shows that the orbit is oriented so that the semi-major axis is pointed toward the observer ($\omega$=270$^{\circ}$) but with some orbital inclination (likely $i$$\sim$41$^{\circ}$) and that \eca\ is in front of its companion at periastron. In our simplistic model the \ion{He}{1} absorption forms in the wind of \eca\ due to ionization by the far-UV flux from \ecb, and that the He$^+$ portion of  \eca's wind approaches the observer prior to periastron passage, consistent with the increasing blueshift seen in the \ion{He}{1} and \ion{H}{1} lines. \\
\indent One surprising result of our  \ion{He}{1} absorption analysis, is the large derived value of the velocity amplitude, $K$=140 \kms. Since the star which is in front at periastron is \eca, the large $K$-value along with the increasingly blueshifted absorption prior to periastron, would imply that \eca\ itself is the lower mass object, or, more precisely, that the star whose wind dominates the system is the lower mass object. Such a scenario is not unusual. In most well-studied colliding wind binaries the lower mass star has the stronger wind. A case in point is WR~140, a canonical WC+O colliding wind system, which, like \ec, has a long period ($P$=7.9 years), high eccentricity \citep[$e$=0.881$\pm$0.005][]{Marchenko03}, hard X-ray emission and in which the weaker-wind O star dominates the mass and luminosity of the system. An exception to this generalization is the LBV+WR system HD~5980, in which the more massive star is an eruptive variable that currently possesses the stronger wind though prior to its eruption in 1994 the more massive star had the weaker wind \citep{Koenigsberger06}.  \\
\indent A more plausible explanation is that the $K$-value derived from the \ion{He}{1} radial velocity curve is overestimated because of ionization effects, as discussed in Section~\ref{rv_i}.  An exact modeling of the effects of \ecb's far-UV flux on the ionization state of \eca's wind depends sensitively on knowledge of \eca's wind structure (which is distorted by the wind-wind interaction) and on the companion's far-UV flux, neither of which is well constrained.  We qualitatively represent these effects, but our model of an undisturbed primary wind being ionized by the far-UV of the secondary is too simple.  Numerous effects may come into play.  For example, the wind from \eca\ piles up on the colliding wind shock, while radiation and thermal energy from the wind shock may also excite helium. As the system approaches periastron, the evidence is strong that the secondary star approaches to within a few AU  of \eca. When this happens the winds and atmospheres of the two stars may experience a major disruption. This could have a major influence on the observed line profiles and measured velocities. The Coriolis effect in this very eccentric orbit greatly influences the structure and therefore, the derived $K$-value. It is a matter of controversy whether the colliding wind interface remains intact at this point.  \\
\indent The \ion{He}{1} and \ion{H}{1} line profiles reveal important information about the binary system. However, only two observations, the first about one week before the minimum and the second one week afterward, measure the rapid changes in velocity in \ion{He}{1} and \ion{H}{1}. Much more frequent observations with high spatial resolution are needed.  We caution observers that because of nebular scatter of the stellar component and contamination of nebular emission lines, sub-arcsecond angular resolution is necessary.

\section{Summary}
We have monitored the \ion{He}{1}  and \ion{H}{1} emission and absorption between 1998.0 and 2004.3. The \ion{He}{1} lines generally show blueshifted peculiar profiles, with a dramatic velocity shift across the spectroscopic minimum. These lines are likely produced by the flux of the hot companion star, though energy released in the wind-wind interface region may also play a role. The \ion{He}{1} lines show a multi-component structure, consisting of a broad P-Cygni feature with narrow peaks superimposed on top. The lines are generally blueshifted for most of the spectroscopic/orbital cycle, showing that they are produced in the portion of the wind of \eca\ which is flowing towards the observer. We have interpreted the narrow emission line components as formed in spatially separated regions in the bow-shock. Hence, the \ion{He}{1} lines forming in \eca's wind in the vicinity of the bow-shock, are the result of the far-UV flux from the hot \ecb. The \ion{H}{1} lines display homogeneous profiles centered at the system velocity and show a smaller shift in velocity across the spectroscopic minimum. The smooth velocity curve for the \ion{H}{1} lines implies line formation from a large part of the system. \\
\indent We have measured the radial velocity and equivalent width variations in the \ion{He}{1} absorption and emission lines and created a radial velocity curve from the absorption lines.  Using this radial velocity curve, we have calculated a mass function for the binary system and estimated the mass of the hidden companion star, and show that either the companion star is the more massive object in the system, or that our estimate of the velocity amplitude is too large due to ionization effects.  We believe that the latter explanation is the more likely one, and that detailed modeling of the radiative transfer in the system may allow us to constrain the mass function more accurately from the \hst\ STIS data. 

 \acknowledgements
The observations were accomplished through STIS GTO, \hst\ GO and \hst\ \ec\ Treasury Team observations and were made with the NASA/ESA \hst\ Science Institute, which is operated by the Association of Universities for Research in Astronomy, Inc., under NASA contract NAS 5-26555. We thank the STScI schedulers who worked closely with T.~R. Gull in designing the observational programs. All planned observations were accomplished at the requested intervals and virtually all observations were optimally exposed. All analysis was done using STIS IDT software tools on reduced data by K.~Davidson, K.~Ishibashi and J.~ Martin made available through the \hst\ \ec\ Treasury public archive.

\bibliographystyle{apj}
\bibliography{ms} 

\end{document}